\documentclass[12pt,letterpaper]{article}
\input{epsf}
\usepackage[pdftex]{graphicx,color}
\usepackage{hyperref}
\input{psfig}
\usepackage{amsmath,amssymb}
\usepackage{fancyhdr}
\newcommand\ltap{\
  \raise.3ex\hbox{$<$\kern-.75em\lower1ex\hbox{$\sim$}}\ }
\newcommand\gtap{\
  \raise.3ex\hbox{$>$\kern-.75em\lower1ex\hbox{$\sim$}}\ }

\newcommand\simge{\mathrel{%
   \rlap{\raise 0.511ex \hbox{$>$}}{\lower 0.511ex \hbox{$\sim$}}}}
\newcommand\simle{\mathrel{
   \rlap{\raise 0.511ex \hbox{$<$}}{\lower 0.511ex \hbox{$\sim$}}}}

\newcommand{\slashchar}[1]%
        {\kern .25em\raise.18ex\hbox{$/$}\kern-.70em #1}
\def\lsim{\mathrel{\raise.3ex\hbox{$<$\kern-.75em\lower1ex\hbox{$\sim$}}}}
\def\gsim{\mathrel{\raise.3ex\hbox{$>$\kern-.75em\lower1ex\hbox{$\sim$}}}}

\newcommand\CL{{\cal L}}
\newcommand\CM{{\cal M}}

\newcommand\CO{{\cal O}}

\newcommand\CZ{{\cal Z}}

\newcommand\ul{\underline}
\newcommand\ol{\overline}
\newcommand\be{\begin{equation}}
\newcommand\ee{\end{equation}}
\newcommand\bea{\begin{eqnarray}}
\newcommand\eea{\end{eqnarray}}
\newcommand\ba{\begin{array}}
\newcommand\ea{\end{array}}
\newcommand\nn{\nonumber}

\newcommand{\half}{\ensuremath{\frac{1}{2}}}

\newcommand{\thalf}{\textstyle{\frac{1}{2}}}
\newcommand{\ttwofour}{\textstyle{\frac{1}{24}}}

\newcommand{\tfourth}{\textstyle{\frac{1}{4}}}

\newcommand{\tsixth}{\textstyle{\frac{1}{6}}}
\newcommand{\teighth}{\textstyle{\frac{1}{8}}}
\newcommand{\ttwelfth}{\textstyle{\frac{1}{12}}}

\newcommand\dagg{\dagger}

\newcommand\leftra{\leftrightarrow}

\newcommand\gev{{\rm GeV}}
\newcommand\tev{{\rm TeV}}

\newcommand\GeV{{\rm GeV}}

\newcommand\ifb{{\rm fb}^{-1}}

\newcommand\ellm{\ell^-}

\newcommand\ellp{\ell^+}

\newcommand\cbeta{c_\beta}
\newcommand\sbeta{s_\beta}

\newcommand{\LGW}{\Lambda_{\rm GW}}

\newcommand{\blam}{\ol{\lambda}}

\newcommand{\bM}{\ol{M}}

\newcommand{\Hpt}{H^{\prime\,2}}

\begin{document}

\title{
\vskip -60mm
\begin{flushright}
  \vskip 6mm
 {\small FERMILAB-PUB-22-667-T\\}
 {\small LAPTH-030/22\\}
 \vskip 5mm
 \end{flushright}
%
 {\Large{\bf The Gildener-Weinberg two-Higgs doublet model at two
     loops}\footnote{This paper is dedicated to Kurt Gottfried and Eric
     Pilon, our friends and collaborators.}}\\
  \medskip } \author{ {\large Estia
    J.~Eichten$^{1}$\thanks{eichten@fnal.gov}}\, and
  Kenneth Lane$^{2}$\thanks{lane@bu.edu}\\
  {\large {$^{1}$}Theoretical Physics Group, Fermi National Accelerator
    Laboratory}\\
  {\large P.O. Box 500, Batavia, Illinois 60510}\\
  {\large $^{2}$Department of Physics, Boston University}\\
  {\large 590 Commonwealth Avenue, Boston, Massachusetts 02215}\\
} \maketitle

\vspace{-1.0cm}

\begin{abstract}

  The Gildener-Weinberg two-Higgs doublet model (GW-2HDM) provides a
  naturally light and aligned Higgs boson, $H = H(125)$. It has been studied
  in the one-loop approximation of its effective potential, $V_1$. An
  important consequence is that the masses of the model's BSM Higgs bosons
  ($H',A,H^\pm$) are bounded by the sum rule
  $\left(M_{H'}^4 + M_A^4 + 2M_{H^\pm}^4\right)^{1/4} = 540\,{\rm
    GeV}$. Although they are well within reach of the LHC, searches for them
  have been stymied by large QCD backgrounds. Another consequence is that $H$
  is highly aligned, i.e., $H$--$H'$ mixing is small and $H$ has only
  Standard Model couplings. A corollary of this alignment is that search
  modes such as $H',\,A \leftra W^+W^-,\,ZZ,\,HZ$ and
  $H^\pm \leftra W^\pm Z,\,W^\pm H$ are greatly suppressed. To assess the
  accuracy of the sum rule and Higgs alignment, we study this model in two
  loops. This calculation is complicated by having many new contributions. We
  present two formulations of it to calculate the $H$--$H'$ mass matrix, its
  eigenvectors $H_1,\,H_2$, and the mass $M_{H_2}$ while fixing
  $M_{H_1}= 125\,\gev$. They give similar results, in accord with the
  one-loop results. Requiring $M_A = M_{H^\pm}$, we find
  $180\,\gev \simle M_{A,H^\pm} \simle 380$--$425\,\gev$ and
  $550$--$700\,\gev \simge M_{H_2} \simge 125\,\gev$, with $M_{H_2}$
  decreasing as $M_{A,H^\pm}$ increase. The corrections to $H$-alignment are
  below $\CO(1\%)$. So, the BSM searches above will remain fruitless. Finding
  the BSM Higgses requires improved sensitivity to their low masses. We
  discuss two possible searches for this.

  \end{abstract}


\newpage

\section*{I. Review and Overview} 

In the Gildener-Weinberg (GW) scheme of electroweak symmetry breaking in
multi-Higgs multiplet models, the scalar potential $V_0(\Phi_i)$ of the
tree-level Lagrangian consists of only quartic interaction
terms~\cite{Gildener:1976ih}. Therefore, so long as all particle masses arise
from the vacuum expectation values (VEVs) $\langle\phi_i\rangle$ of $\Phi_i$,
the theory is classically scale-invariant. This happens if the linear
combination of $\Phi_i$ that is the Higgs boson, $H$, is also a Goldstone
boson of spontaneous breaking of this scale invariance, the dilaton of a flat
minimum of $V_0$ along a ray $0 < \phi < \infty$ in field space.

Because it is such a Goldstone boson, $H$ is the same form of linear
combination of scalars as the Goldstone bosons eaten by $W^\pm$ and $Z$; that
is, in an $N$-multiplet model,
\be\label{eq:Hcombo}
H = \sum_{i=1}^N (\langle\phi_i\rangle/\phi) \phi_i\,
\ee
where $\sqrt{\sum_i \langle\phi_i\rangle^2} = \phi$. This is important:~this
Higgs boson is perfectly aligned, that is, it has exactly the same couplings
to gauge bosons and fermions as the Standard Model (SM)
Higgs~\cite{Boudjema:2001ii,Gunion:2002zf,Carena:2013ooa,Draper:2020tyq,
  Haber:2021zva}. While it is massless at tree-level, the Higgs gets a mass
at the one-loop level of the Coleman-Weinberg effective potential,
$V_1$~\cite{Coleman:1973jx}. The renormalization scale in $V_1$ explicitly
breaks the scale symmetry of $V_0$, inducing a minimum of $V_0 + V_1$ that
picks out a specific value~$v$ of $\phi$. This $v$ is identified as the weak
scale, $246\,\gev$, and it sets the scale of all masses in the
theory.\footnote{For economy of narrative, we are ignoring here the
  spontaneous breaking of the light quarks' chiral symmetry that sets the
  mass scale of the light hadrons.} As we review in this section, the
one-loop corrections to perfect alignment are very small, typically
$\simle \CO(1\%)$ {\em in amplitude}. Thus, the approximate scale symmetry of
GW models makes the Higgs {\em naturally light and
  aligned}~\cite{Lane:2018ycs}.

This naturalness requires no symmetry other than scale invariance. Therefore,
in GW models there are no partners, scalar or fermionic, of the top quark, of
the weak bosons, nor of any other particles except for the additional scalars
occurring in multi-Higgs multiplet models. Nor are there the vectorial
fermions requiring tree-level bare masses. The GW scheme is the only one we
know in which the same agent, the Higgs VEV $v$, is responsible for
electroweak symmetry breaking and for explicit scale symmetry breaking.
Hence, the ``dilaton scale''~$f$ is {\em equal} to
$v$~\cite{Bellazzini:2012vz}. The one sure way to test these models is to
search for the additional Higgs scalars~\cite{Lane:2019dbc,
  Eichten:2021qbm}. They are exceptionally light, with masses below about
$400$--$550\,\gev$ in one-loop order. We review this calculation below in a
two-Higgs-doublet model. This model has three Beyond-Standard-Model Higgs
(BSM) bosons, a $C\!P$-even $H'$, a $C\!P$-odd $A$ and a singly-charged
$H^\pm$ (see the standard Ref.~\cite{Branco:2011iw} for details).

To evaluate the robustness of the one-loop predictions, we extend their
calculation to the two-loop effective potential~\cite{Martin:2001vx}. This is
considerably more complicated than in one loop. Therefore, in Secs.~II
and~III we present two methods of calculating the two-loop contributions to
the $C\!P$-even mass-squared matrix, $\CM^2_{0^+}$, as a function of BSM Higgs
masses $M_A = M_{H^\pm}$ and $M_{H'}$.\footnote{The constraint
  $M_A = M_{H^\pm}$ is motivated by the fact that it makes the contribution
  to the $T$-parameter from the BSM scalars
  vanish~\cite{Battye:2011jj,Pilaftsis:2011ed,Lee:2012jn}.} The two methods
give qualitatively similar results and, so, support the low bound on the
masses of the new Higgs scalars and our earlier conclusion on the degree of
the 125-GeV Higgs boson's alignment. To our knowledge, two-loop calculations
of Gildener-Weinberg multi-Higgs models have not been carried out in this
depth. The experimental consequences of our calculations, including the
impact of ATLAS and CMS searches relevant to the model's BSM Higgs bosons,
are presented in Sec.~IV. Readers interested mainly in these consequences can
skip to Sec.~IV.

The simplest model employing the GW mechanism is the two-Higgs doublet model
(2HDM) proposed by Lee and Pilaftsis in 2012~\cite{Lee:2012jn}. The
tree-level potential of the two doublets is
\bea\label{eq:Vzero}
V_0(\Phi_1,\Phi_2) &=&\lambda_1 (\Phi_1^\dagg \Phi_1)^2 +
\lambda_2 (\Phi_2^\dagg \Phi_2)^2 +
\lambda_3(\Phi_1^\dagg \Phi_1)(\Phi_2^\dagg \Phi_2)\nn \\
&+& \lambda_4(\Phi_1^\dagg \Phi_2)(\Phi_2^\dagg \Phi_1)+
\thalf\lambda_5\left((\Phi_1^\dagg \Phi_2)^2 + (\Phi_2^\dagg
  \Phi_1)^2\right),
\eea
where the doublets are
\be\label{eq:Phii}
\Phi_i = \frac{1}{\sqrt{2}}\left(\ba{c}\sqrt{2} \phi_i^+ \\ \rho_i + i
  a_i \ea\right), \quad i = 1,2,
\ee
and $\rho_i$ and $a_i$ are neutral $C\!P$-even and odd fields. The five
quartic couplings~$\lambda_i$ in Eq.~(\ref{eq:Vzero}) are real and $V_0$ is
$C\!P$-invariant.\footnote{Of course, there is $C\!P$ violation in the CKM
  matrix, but that has negligible effect on our study and we ignore it.}
Positivity of $V_0$ requires that $\lambda_1,\, \lambda_2 > 0$,
This potential is consistent with a $\CZ_2$ symmetry that prevents tree-level
flavor-changing interactions among fermions, $\psi$, induced by neutral
scalar exchange~\cite{Glashow:1976nt}. We define this $\CZ_2$ to be
\be\label{eq:Z2}
\Phi_1 \to -\Phi_1,\,\, \Phi_2 \to \Phi_2, \quad
\psi_L \to -\psi_L,\,\, \psi_{uR} \to \psi_{uR},\,\,
\psi_{dR} \to \psi_{dR}.
\ee
This is the usual type-I 2HDM~\cite{Branco:2011iw}, but with $\Phi_1$ and
$\Phi_2$ interchanged. The net effect of this is that the experimental upper
limit on $\tan\beta = v_2/v_1$ found for this theoretical
model~\cite{Lane:2018ycs} is to be compared to experimental upper limits on
$\cot\beta$ for this and the other three types of 2HDM's with natural flavor
conservation.\footnote{Strictly speaking, in this 2HDM, the VEVs $v_1$ and
  $v_2$ of $\Phi_1$ and $\Phi_2$ have meaning only after scale invariance is
  explicitly broken and $\phi$ in Eq.~(\ref{eq:theray}) has a specific
  value.}
We refer to this model as the GW-2HDM. This type-I coupling was imposed on
the model in 2018 to make it consistent with precision electroweak
measurements at LEP, searches for $t\to H^+b$ at the
Tevatron~\cite{Workman:2022ynf} and the then-current LHC data. The most
stringent constraints came from CMS~\cite{Khachatryan:2015qxa} and
ATLAS~\cite{Aaboud:2018cwk} searches for charged Higgs decay into $t\bar
b$. Consistency with these searches required $\tan\beta \simle 0.50$ for
$180\,\gev < M_{H^\pm} \simle 500\,\gev$. This limit on $\tan\beta$ was
affirmed in Refs.~\cite{Lane:2019dbc, Eichten:2021qbm}.

The trivial minimum of $V_0$ occurs at $\Phi_1 = \Phi_2 = 0$. But a
nontrivial flat minimum of $V_0$ can occur on the ray $0 < \phi < \infty$:
\be\label{eq:theray}
\Phi_{1\beta} = \frac{1}{\sqrt{2}} \left(\ba{c} 0\\ \phi\,\cbeta
  \ea\right),\quad
\Phi_{2\beta} = \frac{1}{\sqrt{2}} \left(\ba{c} 0\\ \phi\,\sbeta \ea\right),
\ee
where $\cbeta = \cos\beta$ and $\sbeta = \sin\beta$ and $\beta \neq 0,\pi/2$
is a fixed angle. The tree-level extremal conditions for this ray are
\bea\label{eq:V0ext}
\left.\frac{\partial V_0}{\partial \rho_1}\right\vert_{\langle \rho_i\rangle}
&=&
\phi^3\cbeta\left(\lambda_1\cbeta^2 + \thalf\lambda_{345}\sbeta^2\right) = 0,
\nn\\
\left.\frac{\partial V_0}{\partial \rho_2}\right\vert_{\langle \rho_i\rangle}
&=&
\phi^3\sbeta\left(\lambda_2\sbeta^2 + \thalf\lambda_{345}\cbeta^2\right) = 0,
\eea
where $\lambda_{345} = \lambda_3 + \lambda_4 + \lambda_5$.  It can be proved
that $V_0(\Phi_{i\,\beta}) = 0$ and, in fact, that any such purely quartic
potential as well as its first derivative vanish at any
extremum~\cite{Lane:2019dbc}. These conditions on the quartic couplings,
\be\label{eq:lamconds}
\lambda_1 = -\thalf\lambda_{345} \tan^2\beta, \qquad
\lambda_2 = -\thalf\lambda_{345} \cot^2\beta,
\ee
remain true and in force {\em in all orders} of the loop expansion for the
effective potential~\cite{Gildener:1976ih}. This will be important in our
subsequent development.

The eigenvectors and eigenvalues of the scalars' squared ``mass'' matrices in
tree approximation are given by
\bea\label{eq:mevec}
\left(\ba{c} z \\ A\ea\right) &=& \left(\ba{cc} \cbeta & \sbeta\\
     -\sbeta & \cbeta\ea\right) \left(\ba{c} a_1 \\ a_2\ea\right),
\quad M_z^2 = 0, \,\,\, M_A^2  = -\lambda_5 \phi^2;\nn\\
%
\left(\ba{c} w^\pm \\ H^\pm\ea\right) &=& \left(\ba{cc} \cbeta & \sbeta\\
     -\sbeta & \cbeta\ea\right) \left(\ba{c} \phi_1^\pm \\ \phi_2^\pm\ea\right),
\quad M_{w^\pm}^2 = 0, \,\,\, M_{H^\pm}^2 = -\thalf\lambda_{45}
\phi^2;\nn \\
%
\left(\ba{c} H \\ H'\ea\right) &=& \left(\ba{cc} \cbeta & \sbeta\\
     -\sbeta & \cbeta\ea\right) \left(\ba{c} \rho_1 \\ \rho_2\ea\right),
\quad M^2_H = 0, \,\,\, M^2_{H'}  = -\lambda_{345} \phi^2.
\eea
It is important to note that the extremal conditions~(\ref{eq:V0ext}) are
equivalent to the vanishing of the Goldstone boson masses, $M_z$ and
$M_{w^\pm}$. The ray~(\ref{eq:theray}) is a (flat) minimum, with $V_0 = 0$,
so long as the $M^2$ are non-negative, i.e., that $\lambda_5$,
$\lambda_{45} = \lambda_4 + \lambda_5$ and $\lambda_{345}$ are negative.
The $C\!P$-even scalar $H$ is the dilaton and, as discussed above, it is the
same linear combination of fields as $z$ and $w^\pm$ are; i.e., $H$ is
aligned.  Alignment will be modified in higher orders, but only slightly.

At this point and for our discussion of this model beyond the tree
approximation, it is convenient to use the ``aligned basis'' of the Higgs
fields because, in the GW-2HDM, $H$ is very nearly aligned and separated from
the BSM Higgs fields $H',A,H^\pm$ {\em through two-loop order} in this
basis.\footnote{It is also called the Higgs basis; see
  Ref.~\cite{Branco:2011iw} and references therein.} The aligned basis is:
\bea\label{eq:aligned}
\Phi &=& \Phi_1\cbeta + \Phi_2\sbeta = \frac{1}{\sqrt{2}} \left(\ba{c}
  \sqrt{2}w^+\\ H + iz\ea\right), \nn\\\\
\Phi' &=& -\Phi_1\sbeta  + \Phi_2\cbeta = \frac{1}{\sqrt{2}} \left(\ba{c}
  \sqrt{2}H^+\\ H' + iA \ea\right).\nn
\eea
On the ray Eq.~(\ref{eq:theray}) on which $V_0$ has nontrivial extrema, these
fields are
\be\label{eq:alignedray}
\Phi_{\beta} = \frac{1}{\sqrt{2}} \left(\ba{c} 0\\ \phi \ea\right),\quad
\Phi'_{\beta} = \frac{1}{\sqrt{2}} \left(\ba{c} 0\\ 0 \ea\right),
\ee
The tree-level extremal conditions in this basis are
\bea
\label{eq:tree}
&& \left.\frac{\partial V_0}{\partial H}\right\vert_{\langle\,\rangle} =
\phi^3\left[\lambda_1\cbeta^4 + \lambda_2\sbeta^4 +
  \lambda_{345}\sbeta^2\cbeta^2\right] = 0,\nn\\\\
&&
\left.\frac{\partial V_0}{\partial H'}\right\vert_{\langle\,\rangle} =
\thalf\phi^3\left[(2\lambda_2\sbeta^2 +\lambda_{345}\cbeta^2) -
  (2\lambda_1\cbeta^2 +\lambda_{345}\sbeta^2)\right]\sbeta\cbeta = 0,\nn
\eea
where $\langle\,\rangle$ means that the derivatives are evaluated at
$\langle H\rangle = \phi$ while $\langle H'\rangle$ and all other VEVs equal
zero. Using Eqs.~(\ref{eq:tree}), the tree potential is\footnote{Note that
  that there are no higher powers of $H,z,w^\pm$ than quadratic in
  Eq.~(\ref{eq:V0align}).}
\bea
\label{eq:V0align}
V_0 &=& -2\lambda_{345}\left[{\thalf}\left(\Phi^\dagg\Phi'+
  \Phi^{\prime\,\dagg}\Phi\right) + \Phi^{\prime\,\dagg}\Phi'
\cot 2\beta\right]^2 \nn\\
&&-\lambda_{45}\left[\left(\Phi^\dagg\,\Phi\right)
  \left(\Phi^{\prime\,\dagg}\,\Phi'\right) -
\left(\Phi^\dagg\,\Phi'\right) \left(\Phi^{\prime\,\dagg}\Phi\right)
\right]
+{\thalf} \lambda_5\left[\Phi^\dagg\Phi' -  \Phi^{\prime\,\dagg}\Phi\right]^2
\\
\nn\\
\label{eq:Vzalign}
&=& -{\thalf}\lambda_{345}\left[HH'+zA+ w^+H^- + H^+w^-
    + \left(\Hpt+A^2+2H^+H^-\right)\cot 2\beta\right]^2\nn\\
&& - {\thalf}\lambda_{45}\bigl[\left(H^2 + z^2\right)H^+H^- + \left(\Hpt+
  A^2\right)w^+w^-\nn\\
&&\quad -(HH'+ zA)(w^+H^- + H^+w^-) -i(HA-zH')(w^+H^- - H^+w^-)\bigr]\nn\\
&& - {\thalf}\lambda_5 \left[HA-zH' +i(w^+H^- - H^+w^-)\right]^2.
\eea
The form of Eq.~(\ref{eq:Vzalign}) will be used in Sec.~II to define the
mass-dependent scalar couplings that appear in the two-loop calculations.
The tree-level ``mass'' matrices of the Higgs bosons are\footnote{The quotes
  around ``mass'' are there because $0 < \phi < \infty$.}
\bea
\label{eq:mevecA}
\CM^2_{0^-} &=& \left(\ba{cc} 0 & 0\\ 0 & M^2_A\ea\right) \qquad
{\text{with\,\,}} M^2_A  = -\lambda_5 \phi^2;\\
\label{eq:mevecpm}
\CM^2_{\pm} &=& \left(\ba{cc} 0 & 0\\ 0 & M^2_{H^\pm}\ea\right) \,\,\quad
{\text{with\,}}  M^2_{H^\pm}  = -\thalf\lambda_{45} \phi^2;\\
\label{eq:mevecH}
\CM^2_{0^+} &=& \left(\ba{cc} 0 & 0\\ 0 & M^2_{H'}\ea\right) \,\,\,\quad
{\text{with\,}}  M^2_{H'}  = -\lambda_{345} \phi^2.
\eea

The Coleman-Weinberg effective potential in one-loop order is a sum over the
heavy particles in the model~\cite{Jackiw:1974cv,Martin:2001vx,Lee:2012jn}:
\be\label{eq:Vone} V_1 = \frac{1}{64\pi^2}\sum_n \alpha_n\bM_n^4
\left(\ln\frac{\bM_n^2}{\LGW^2} - k_n\right).
\ee
For $n = (W^\pm,Z,t_L+t_R^c, H',A,H^\pm)$, $\alpha_n = (6,3,-12,1,1,2)$
counts the degrees of freedom of particle~$n$ and $k_n = 5/6$ for the weak
gauge bosons and~3/2 for the scalars and the top-quark Weyl
fermions.\footnote{$V_1$ is calculated in the Landau gauge using the
  ${\ol{\rm MS}}$ renormalization scheme.}  The background-field dependent
masses $\bM_n^2$ in Eq.~(\ref{eq:Vone}) are~\cite{Jackiw:1974cv,Lee:2012jn}
\be\label{eq:bkgM}
\bM_n^2 = \left\{\ba{l} M_n^2\left(2 \left(\Phi^\dagg \Phi +
      \Phi^{\prime\,\dagg} \Phi'\right)/\phi^2\right) =
   M_n^2 \left((H^2 + H^{\prime\,2} + \cdots)/\phi^2\right), \qquad n \neq t, b\\
    M_t^2 \left(2\Phi_1^\dagg\Phi_1/(\phi\cbeta)^2\right) =
    M_t^2\left((H - H'\tan\beta)^2 + \cdots\right)/(\phi)^2  \ea\right.,
\ee
where $M_n^2 \propto \phi^2$ is the actual squared mass of particle~$n$ at
scale~$\phi$. (We put $M_b^2= 0$ in calculations.) The form of $\bM_t^2$ is
dictated by the type-I coupling of fermions to the $\Phi_1$ doublet in
Eq.~(\ref{eq:Z2}).\footnote{To avoid the confusion of too much notation, we
  use the same symbol, e.g. $H$, for the quantum field of particle~$H$ and
  for its classical counterpart in the field-dependent masses. Context will
  dictate which field is being used. However, for clarity in the
  field-dependent cubic couplings introduced in Sec.~IIb, we denote the
  classical counterpart of field $H$ by $H_c$, etc.}  Finally, $\LGW$ is a
renormalization scale that will be fixed relative to the Higgs VEV
$v = 246\,\gev$ in Eqs.~(\ref{eq:lgw2}--\ref{eq:lgw3}) below.

Following GW~\cite{Gildener:1976ih}, extremal conditions and masses are
obtained by evaluating derivatives of the effective potential
$V_{\rm eff} = V_0 + V_1 + V_2 + \cdots$ at $\langle\,\rangle +$ possible
shifts $\delta H$ and $\delta H'$ in the VEVs of $H$ and $H'$.\footnote{The
  VEVs of the mass eigenstate Higgs bosons, called $H_1$ and $H_2$ in
  Eq.~(\ref{eq:H1H2evecs}), will be fixed to
  $\langle H_1\rangle^2 + \langle H_2\rangle^2 = v^2 =
  (246.2\,\gev)^2$. Also see Eq.~(\ref{eq:vdef}) and the accompanying
  footnote.}  We assume that these shifts have a loop expansion, e.g.,
$\delta H' = \delta_1H' + \delta_2H' + \cdots.$ The extremal conditions at
one-loop order are~\cite{Gildener:1976ih}
\bea
\label{eq:extrxa}
&&\left.\frac{\partial(V_0+V_1)}{\partial
    H}\right\vert_{\langle\,\rangle + \delta_1H + \delta_1H'}
= 0,\\
\label{eq:extrxb}
&&\left.\frac{\partial (V_0+V_1)}{\partial
    H'}\right\vert_{\langle\,\rangle  + \delta_1H + \delta_1H'} = 0.
\eea
Expanding Eqs.~(\ref{eq:extrxa},\ref{eq:extrxb}) to $\CO(V_1)$
and using Eq.~(\ref{eq:mevecH}), these conditions become
\bea
\label{eq:extrx1}
&& \left.\frac{\partial V_1}{\partial H}\right\vert_{\langle\,H\rangle = v} =
\frac{1}{16\pi^2 v} \sum_{n}\alpha_n
M_n^4\left(\ln\frac{M_n^2}{\LGW^2} +\half - k_n\right) = 0,\hspace{1.0cm}\\
\label{eq:extrx2}
&& M^2_{H'}\,\delta_1H' -\frac{\alpha_t M_t^4\tan\beta}{16\pi^2 v}
\left(\ln\frac{M_t^2}{\LGW^2} +\half - k_t\right) = 0,\hspace{1.0cm}
\eea
where the derivative with respect to $H'$ of the $n\neq t$ terms in $V_1$
vanishes because those terms are quadratic in $H'$. Thus,
\be\label{eq:del1Hpr}
\delta_1H'=  -\frac{1}{M^2_{H'}}\left.\frac{\partial V_1}{\partial
    H'}\right\vert_{\langle\,\rangle} =
 \frac{\alpha_t M_t^4\tan\beta}{16\pi^2 M^2_{H'}v}
 \left(\ln\frac{M_t^2}{\LGW^2} +\half - k_t\right),
\ee
the typical tadpole result~\cite{Coleman:1963pj,Coleman:2011xi}. Also,
because $\delta_1H$ is not determined in $\CO(V_1)$, we are free to set
it. We expect from Eq.~(\ref{eq:tandelta}) below that
$\delta_1H = \CO(\delta_1H'\times \delta_1) = \CO(V_2)$, where $\delta_1$ is
the one-loop-induced $H$--$H'$ mixing angle; therefore, we set
\be\label{eq:del1H}
\delta_1H = 0.
\ee

A particular scale~$\phi = v$ appears in Eqs.~(\ref{eq:extrx1},
\ref{eq:extrx2}) because, for nontrivial extrema with $\beta \neq 0,\,\pi/2$,
a deeper minimum than the {\em vanishing} zeroth-order ones {\em can} appear
there:
$(V_0 + V_1)_{\langle H\rangle = v} < V_{0\beta} = V_0(0) + V_1(0) = 0$.  In
that case, Eq.~(\ref{eq:extrx1}) is equivalent to a relation between the
renormalization scale $\LGW$ and the Higgs VEV~$v$:
\be\label{eq:lgw2}
\ln\left(\frac{\LGW^2}{v^2}\right) = \frac{A}{B} + \half,
\ee
%
%
where
\be\label{eq:lgw3}
A = \sum_n \alpha_n M_n^4 \left(\ln\frac{M_n^2}{v^2} - k_n\right),\quad
B = \sum_n \alpha_n M_n^4.
\ee
At $\langle\,\rangle$, $\bM^2_n = M^2_n$, and the effective potential is
\be\label{eq:V1_extr}
\left.(V_0+V_1)\right\vert_{\langle\,\rangle} =
  \frac{1}{64\pi^2}\left(A + B\ln\frac{v^2}{\LGW^2}\right) = -\frac{B}{128\pi^2}.
\ee
Thus, unless $B > 0$, this extremum cannot be a minimum because otherwise it
has no finite bottom for $v \to \infty$~\cite{Gildener:1976ih}.  Despite the
large negative top-quark term in $B$, the contribution of the extra Higgs
bosons can make it positive.  With the minimum occurring at the particular
value~$\phi = v$, the scale invariance of the tree approximation is now
explicitly broken and the Higgs boson $H$ gets a nonzero mass. Note that all
$M_n^2 \propto v^2$ so that the right side of Eq.~(\ref{eq:lgw2}) is a
function of only gauge, Higgs-boson and the top-quark Yukawa couplings. The
VEVs of $\Phi_1$ and $\Phi_2$ are $v_1 = v\cos\beta$ and $v_2 = v\sin\beta$,
with $\tan\beta = v_2/v_1$ as usual in a 2HDM.

\begin{figure}[t!]
 \begin{center}
\includegraphics[width=2.65in,
height=2.65in]{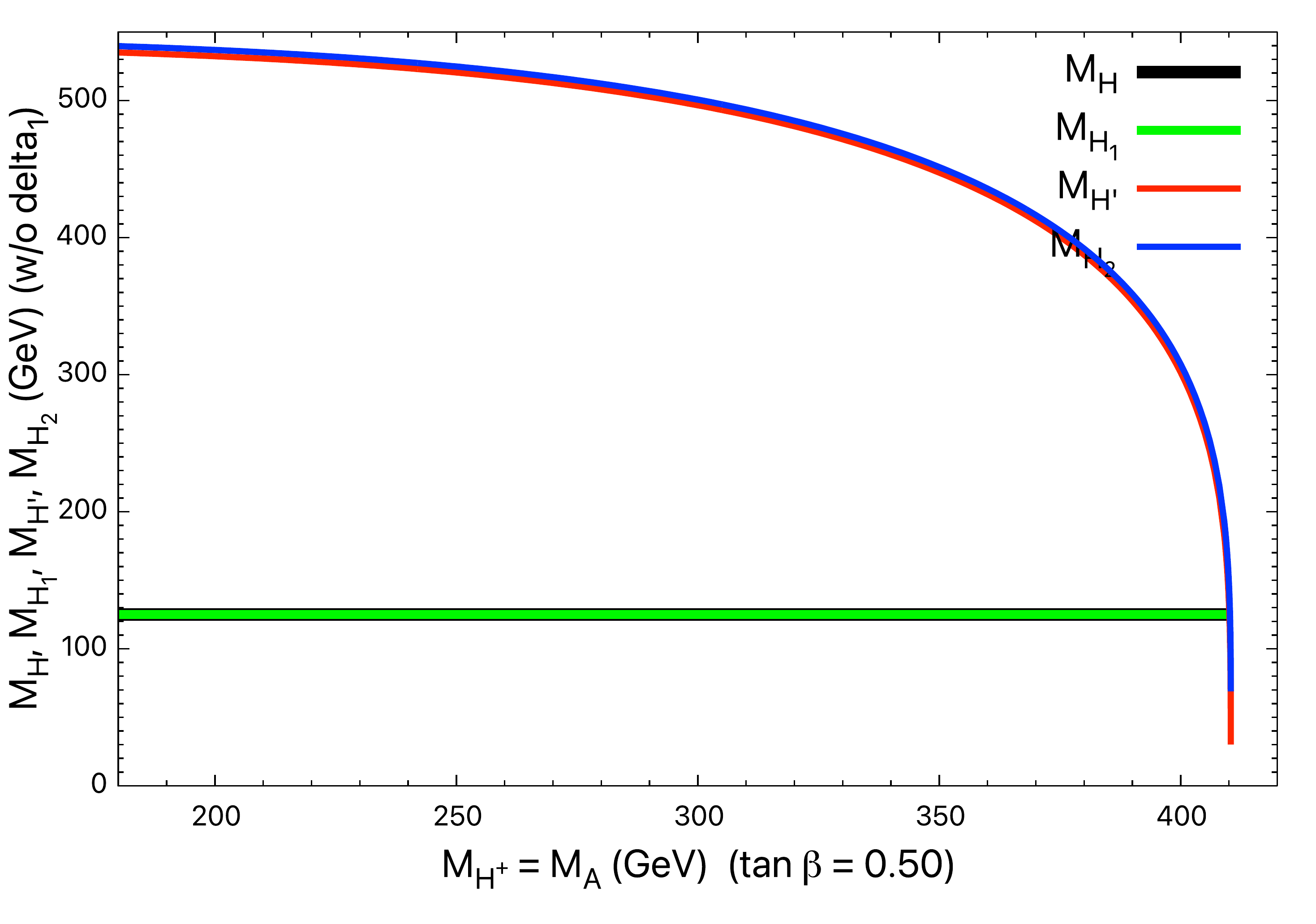}
\includegraphics[width=2.65in,
height=2.65in]{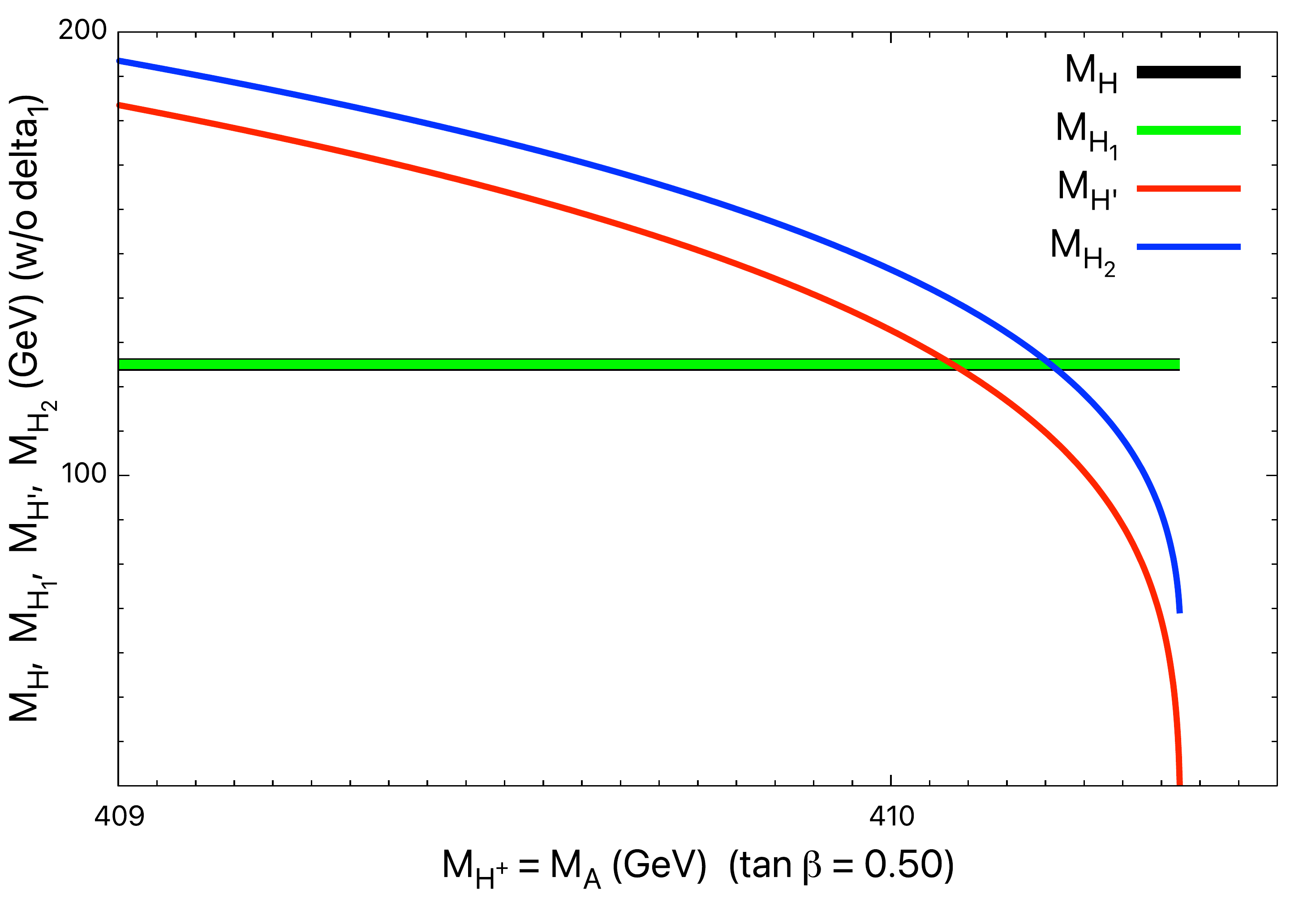}
\caption{Left: The $C\!P$-even Higgs masses: $M_H = 125\,\gev$ in
  Eq.~(\ref{eq:MHsq}), $M_{H'}$ from the sum rule Eq.~(\ref{eq:MHsum}), and
  the eigenvalues $M_{H_1}$ and $M_{H_2}$ from Eq.~(\ref{eq:H1H2evals}) in
  the strict one-loop approximation. The masses are plotted
  vs.~$M_A = M_H^\pm$ from $180\,\gev$ to $410.5\,\gev$ where $M_{H'}$ is
  rapidly approaching zero. Here, $\tan\beta = 0.50$~\cite{Lane:2018ycs};
  only small one-loop masses are sensitive to that choice. Right: A close-up
  of the endpoint of the tree-level and one-loop masses of the $C\!P$-even
  Higgs bosons.}
  \label{fig:V1masses}
 \end{center}
 \end{figure}

\begin{figure}[t!]
  \begin{center}
\includegraphics[width=2.65in,
height=2.65in]{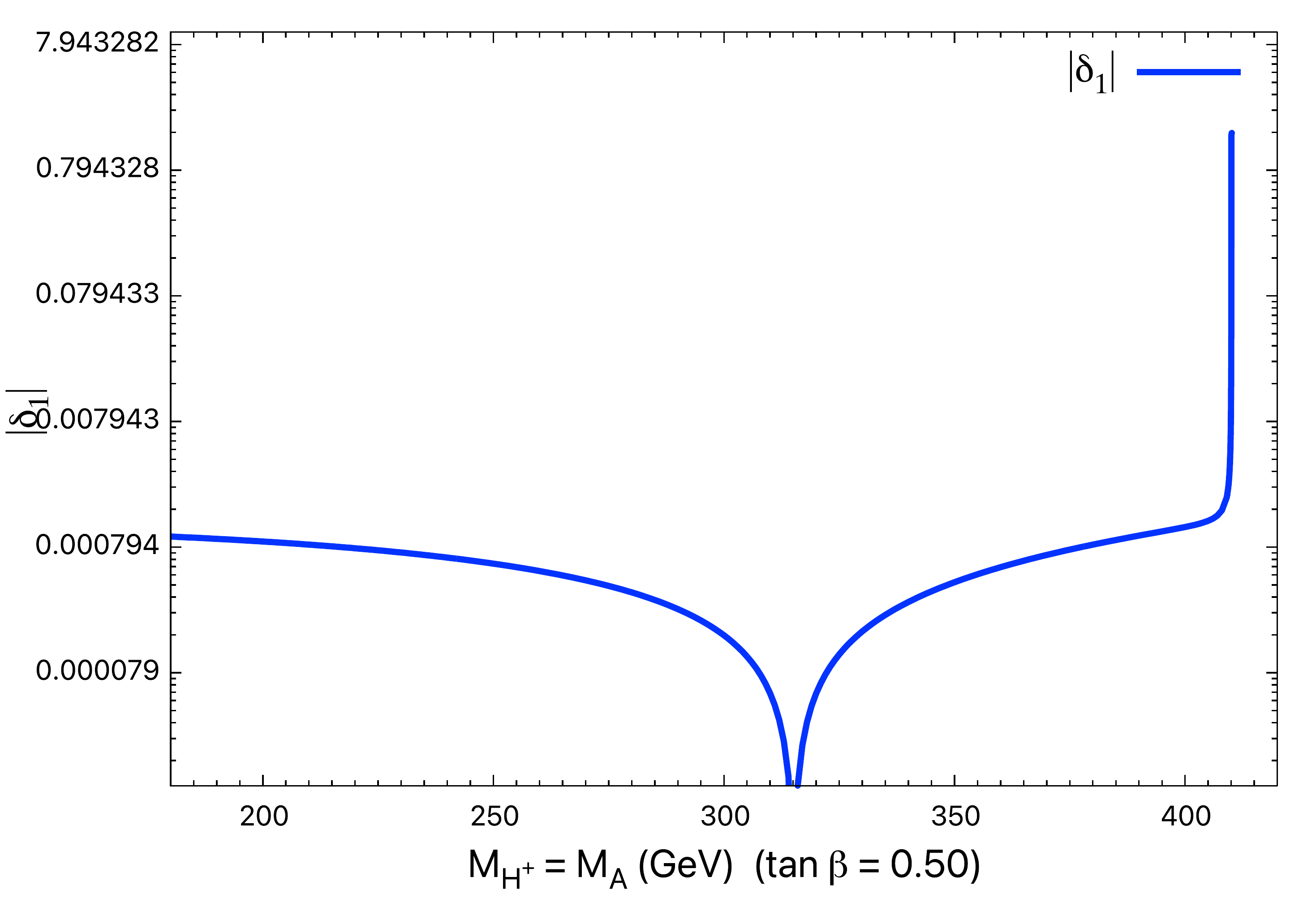}
\includegraphics[width=2.65in,
height=2.65in]{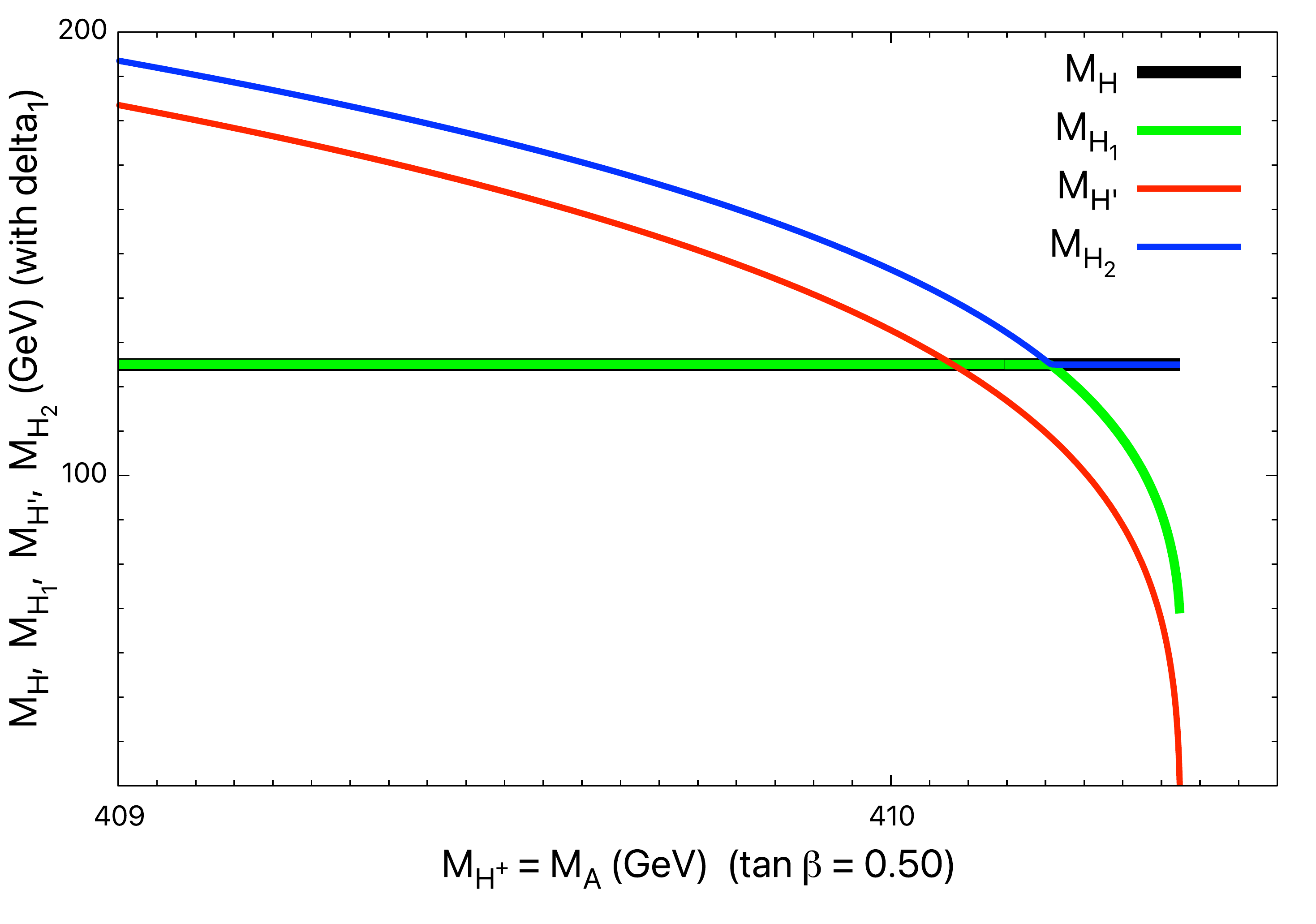}
\caption{The magnitude of the one-loop $H$--$H'$ mixing angle $\delta_1$ (in
  radians) vs.~$M_A = M_{H^\pm}$ at $\tan\beta= 0.50$. Note that
  $\CM^2_{HH'}, \,\delta_1 \propto \tan\beta.$ Left: The full range from
  $M_{H^\pm,A} = 180\,\gev$ to the sum rule cutoff at $410.5\,\gev$. Below
  $M_{H^\pm,A} = 315\,\gev$ the numerator $\CM^2_{HH'}$ of $\tan 2\delta_1$
  is negative and so is $\delta_1$; $\CM^2_{HH'}$ and $\delta_1$ change sign
  at $M_{H^\pm,A} = 315\,\gev$.  Right: A close-up of masses at the endpoint
  as $M_{H'} \to 0$, showing the level repulsion at $M_A = 410.2\,\gev$
  between $M_{H_1}$ and $M_{H_2}$. There, $M_{H_1} \to M_{H'}$ and
  $M_{H_2} \to M_H = 125\,\gev$. In this region $\delta_1 \ge \pi/ 4$ and the
  validity of the loop perturbation expansion has broken down.}
\label{fig:V1angle}
 \end{center}
 \end{figure}

The $C\!P$-even Higgs mass matrix to $\CO(V_1)$ in the aligned basis is
\be\label{eq:M0sq}
\CM^2_{0^+} =
\left(\ba{cc} \left(\partial^2 V_1/\partial H^2\right) &
  \left(\partial^2 (V_0+V_1)/\partial H\partial H'\right) \\
  \left(\partial^2 (V_0+V_1)/\partial H\partial H'\right) &
  \left(\partial^2(V_0+V_1)/\partial
     H^{\prime\,2}\right) \ea\right)_{{\langle\,\rangle} + \delta_1H'},
\ee
where, again using Eq.~(\ref{eq:extrx1}),
\bea
\label{eq:M0sq1}
\CM^2_{HH} &=& \left.\frac{\partial^2 V_1}{\partial
    H^2}\right\vert_{\langle\,\rangle} =
\frac{1}{8\pi^2 v^2}\sum_n \alpha_n M_n^4 \equiv \frac{B}{8\pi^2 v^2},\\
%
\label{eq:M0sq2}
\CM^2_{HH'} &=& \left.\frac{\partial^3 V_0}{\partial H\partial
    \Hpt}\right\vert_{\langle\,\rangle}\delta_1H' +
    \left.\frac{\partial^2 V_1}{\partial H\partial
        H'}\right\vert_{\langle\,\rangle}\nn\\
&=& \frac{2M_{H'}^2\,\delta_1H'}{v}
- \frac{3\alpha_t M_t^4\tan\beta}{16\pi^2v^2}\left(\ln\frac{M_t^4}{\LGW^2}
+ \frac{7}{6} - k_t\right)\nn\\
&=& -\frac{\alpha_t M_t^4\tan\beta}{16\pi^2 v^2} \left(\ln
  \frac{M_t^2}{\LGW^2} + \frac{5}{2} - k_t\right), \\
\label{eq:M0sq3}
\CM^2_{H'H'} &=& \left.\frac{\partial^2 V_0}{\partial
    H^{\prime\,2}}\right\vert_{\langle\,\rangle} +
                 \left.\frac{\partial^3 V_0}{\partial
    H^{\prime\,3}}\right\vert_{\langle\,\rangle}\delta_1H' +
\left.\frac{\partial^2 V_1}{\partial
    H^{\prime\,2}}\right\vert_{\langle\,\rangle} \nn\\
&=& M_{H'}^2 + \frac{6 M_{H'}^2\cot 2\beta\,\delta_1H'}{v} \nn\\
&+& \frac{\alpha_t M_t^4(3\tan^2\beta - 1)}{16\pi^2 v^2}
\left(\ln\frac{M_t^2}{\LGW^2} + \frac{1}{2} - k_t\right)
+ \frac{2\alpha_t M_t^4\tan^2\beta}{16\pi^2 v^2}\nn\\
&=& M^2_{H'} + \frac{\alpha_t M_t^4}{8\pi^2 v^2}\left(\ln\frac{M_t^2}{\LGW^2} +
  \frac{1}{2} - k_t + \tan^2\beta\right).
%
\eea
The eigenvectors $H_1$, $H_2$ and eigenvalues of $\CM^2_{0^+}$, with
$M^2_{H_1} < M^2_{H_2}$, are
\bea
\label{eq:H1H2evecs}
H_1 &=& H\cos\delta_1 - H'\sin\delta_1,\nn\\
H_2 &=& H\sin\delta_1 + H'\cos\delta_1,\\
\label{eq:H1H2evals}
M^2_{H_1} &=&
\CM^2_{HH}\cos^2\delta_1+\CM^2_{H'H'}\sin^2\delta_1 - 2\CM^2_{HH'}\sin\delta_1
\cos\delta_1,\nn\\
M^2_{H_2} &=&
\CM^2_{HH}\sin^2\delta_1+\CM^2_{H'H'}\cos^2\delta_1 + 2\CM^2_{HH'}\sin\delta_1
\cos\delta_1,
\eea
where $\delta_1$ is the $H$--$H'$ mixing angle~$\delta$ in the
{\em one-loop} approximation to
\be\label{eq:tandelta}
\tan 2\delta = \frac{2\CM^2_{HH'}}{\CM^2_{H'H'} - \CM^2_{HH}} \cong
-\frac{\alpha_t M^4_t \tan\beta}{8\pi^2 v^2 M^2_{H'}}
\left(\ln\frac{M_t^2}{\LGW^2} + \frac{5}{2} - k_t\right) + \CO(V_2).
\ee
These eigenmasses and the angle $\delta_1$ are displayed for the GW-2HDM in
Figs.~\ref{fig:V1masses}, \ref{fig:V1angle} and will be discussed below.

The one-loop GW-2HDM formula for the Higgs boson's mass, $M_H = 125\,\gev$,
is
%
\be\label{eq:MHsq}
 M^2_H = \frac{B}{8\pi^2 v^2} + \CO(V_2) = \frac{1}{8\pi^2 v^2}\left(6M_W^4 +
   3M_Z^4 + M_{H'}^4 + M_A^4 + 2M_{H^\pm}^4 - 12m_t^4\right).
\ee
Thus, $B$ is positive, as required so that
$\left.(V_0 + V_1)\right\vert_{\langle\,\rangle} < 0$.  This constrains the
BSM Higgs masses and implies a simple and important sum rule on them:
\be\label{eq:MHsum}
\left(M_{H'}^4 + M_A^4 + 2M_{H^\pm}^4\right)^{1/4} = 540\,\gev.
\ee
This sum rule holds in the one-loop approximation of {\em {\ul{any}}} GW
model of electroweak symmetry breaking in which the only weak bosons are $W$
and $Z$ and the only heavy fermion is the top quark. Thus, the larger the
Higgs sector, the lighter will be the masses of at least some of the BSM
Higgs bosons expected in a GW model. Its importance is that these models
predict extra Higgs bosons at surprisingly low masses. In the GW-2HDM, they
have conventional decay modes, discussed at length in
Refs.~\cite{Lane:2018ycs,Lane:2019dbc,Eichten:2021qbm}. Determining the sum
rule's reliability is a main motivation for extending the calculation of
$\CM^2_{0^+}$ to two loops.

Equations~(\ref{eq:del1Hpr}) and (\ref{eq:M0sq1})--(\ref{eq:tandelta}) a
establish a connection between the top quark and Higgs alignment: If it
were not for the Glashow-Weinberg constraint on the Higgs couplings to
quarks~\cite{Glashow:1976nt} and the top quark's large mass (hence its
appearance in $V_1$), $\delta_1H'$ and $\delta_1$ would vanish and
$\CM^2_{0^+}$ would be diagonal~\cite{Eichten:2021qbm}. This degree of Higgs
alignment means that standard techniques of searching for the BSM Higgs
bosons $H',A$ and $H^\pm$ via their couplings to $W^+W^-$, $ZZ$ and
$W^\pm Z$, both in fusion production and decay and in $H',A \to ZH$ and
$H^\pm \to HW^\pm$, will continue to come up empty-handed; see
{\url{https://twiki.cern.ch/twiki/bin/view/AtlasPublic}} and
{\url{https://cms-results.web.cern.ch/cms-results/public-results/publications/HIG/SUS.html}},
and the electroweak couplings of the GW-2HDM scalars in Sec.~IV,
Eq.~(\ref{eq:PhiEW}). The rates for these processes are proportional to
$\delta_1^2 \sim (\delta_1H'/v)^2 \simle
10^{-3}$~\cite{Lane:2018ycs,Lane:2019dbc}. An equivalent consequence of the
top-quark's connection to alignment is that it is responsible for the
one-loop VEV $\delta_1H'$ acquired by the other $C\!P$-even Higgs, $H'$.

The nearly diagonal nature of $\CM^2_{0^+}$ --- that
$M^2_{H_1} \equiv M^2_H \cong M^2_{HH}$ and $M^2_{H_2} \cong M^2_{H'H'}$ ---
is illustrated in Fig.~\ref{fig:V1masses} where the mass pairs are plotted
versus $M_A = M_{H^\pm}$. In the left panel, where
$180\,\gev \le M_{H^\pm,A} < 410.5\,\gev$, the masses in each pair appear to
be on top of other each other. As the sum rule~(\ref{eq:MHsum}) forces
$M_{H'} \to 0$ at $M_A = M_{H^\pm} = 410.5\,\gev$, the difference in the $H'$
mass pairs due to the top-quark term in $\CM^2_{H'H'}$ is seen in the right
panel.

 Examples of how unimportant the top-quark terms are, except for small
 $M_{H'}$, are displayed in Table~\ref{tab:tops}. Note how sensitive $M_{H'}$
 and the eigenvalues $M_{H_2}$ are as the endpoint of the sum
 rule~(\ref{eq:MHsum}) is approached.
\begin{table}[!h]
  \begin{center}{
  \begin{tabular}{|c|c|c|c|c|}
  \hline
$M_A = M_{H^\pm}$ & $M_{H'}$ & $\delta_1$ & $\delta_1H'$ &$M_{H_2}$ \\
\hline\hline
375.0 & 400.6 & $0.76\times 10^{-3}$  & 1.55 & 404.9\\
409.8 & 147.6 & $0.31\times 10^{-2}$  & 12.3 & 160.0 \\
410.21 & 108.7 &$0.603 \times 10^{-1}$ & 22.9 & 125.2  \\
\hline
\end{tabular}}
\caption{Examples of the approach to the breakdown of the validity of the
  one-loop expansion as the endpoint $M_A = 410.5\,\gev$ of the sum
  rule~(\ref{eq:MHsum}) is approached. Masses are in GeV and
  $\tan\beta = 0.50$.}
\label{tab:tops}
\end{center}
\end{table}

 Reference~\cite{Lee:2012jn} demonstrated a level repulsion between $M_{H_1}$
 and $M_{H_2}$ as $M_{H'} \to 0$. We can reproduce that here by using the
 full Eq~(\ref{eq:H1H2evals}) with $\tan\delta_1$ given by using all
 $\CO(V_1)$ terms in the first equality of Eq.~(\ref{eq:tandelta}). This
 gives contributions of $\CO(V_2)$ which become appreciable to the
 eigenmasses when $M_{H'} \to 0$. We illustrate this in
 Fig.~\ref{fig:V1angle}. In the left panel the angle~$|\delta_1|$ is plotted
 vs.~$M_A$. Below $M_A \simeq 380\,\gev$ the angle is very small,
 $|\delta_1|\simle 10^{-3}$ and it changed sign from negative to positive at
 $M_A = 315\,\gev$. Above $M_{A} \simeq 380\,\gev$, the sum rule starts to
 force $M_{H'} \to 0$, the denominator $\CM^2_{H'H'}-\CM^2_{HH}$ in
 $\tan\delta_1$ decreases rapidly above $M_{H^\pm,A} = 410\,\gev$, changing
 sign at $410.14\,\gev$. Consequently, $|\delta_1|$ rises rapidly from
 $\sim 10^{-3}$, passing through $\pi/4$ on its way to $\pi/2$ when
 $M_{H'} \to 0$. Here, this excursion of the mixing angle is the signal
 of level repulsion, clearly seen in the right panel. The magnitude of the
 angle $\delta_1$ and the swapping of the two $C\!P$-even levels in this
 region signal the breakdown of the validity of the loop perturbation
 expansion.~\footnote{The explanation for this phenomenon in
   Ref.~\cite{Lane:2018ycs} was incorrect also, but for a different reason.}

 In Sec.~IIa we present a formalism for calculating the extremal conditions
 and the $C\!P$-even masses of the two-loop effective potential,
 $V_{\rm eff} = V_0 +V_1 + V_2$, of the GW-2HDM model. This formalism is the
 straightforward generalization to two loops of that in
 Ref.~\cite{Gildener:1976ih}. Still working in the aligned basis, we expand
 derivatives of $V_{\rm eff}$ about their zeroth-order VEVs
 (Eq.~(\ref{eq:alignedray})) allowing for shifts in the VEV's of $H$ and $H'$
 while keeping their RMS equal to~$v$ (see Eq.~(\ref{eq:vdef})). In these
 calculations, we keep terms of at most $\CO(V_2)$, discarding those that are
 formally of higher order in the loop expansion.  We call this procedure the
 ``perturbative method.''

 In Sec.~IIb we simplify our calculation considerably by keeping only the
 all-Higgs-scalar terms in $V_2$. This is quite a good approximation for this
 method; see Fig.~\ref{fig:B1ratios}. In this section we follow
 Martin~\cite{Martin:2001vx} and define the field-dependent triple-scalar
 couplings needed for these calculations.

 Even in this approximation, the two-loop generalization of
 Eqs.~(\ref{eq:MHsq}, \ref{eq:MHsum} is intractable, so we must resort to a
 purely numerical scheme to determine the BSM scalar masses in terms of
 $M_{H_1}$ and $M_W$, $M_Z$, $M_t$. This is done in Secs.~IIb,c. The basis of
 this scheme is that, to $\CO(V_2)$, the $C\!P$-even mass-squared matrix
 $\CM^2_{0^+}$ has positive eigenvalues with $M^2_{H_1}$ close to
 $(125\,\gev)^2$. For this, the two-loop extremal conditions are used to
 determine the corrections to $\LGW$ and the shifts $\delta_2H$, $\delta_2H'$
 in the $C\!P$-even Higgs VEVs. This procedure does not guarantee that
 $\det(\CM^2_{0^+}) > 0$ for the allowed range of $M_A = M_{H^\pm}$ and
 $M_{H'}$ as it does in $\CO(V_1)$. This determinant has terms of $\CO(V_4)$,
 coming from the square of $\CM^2_{HH'}$, for example, and they are not
 small. But it does not contain all fourth-order terms; others will come from
 the three- and four-loop effective potential.\footnote{This seems to be a
   problem with no end unless successive loop contributions become negligibly
   small. Another facet of this will occur in Sec.~III.}

 We find two ``branches'', $B1$ and $B2$, of the $H_2$ mass for
 $M_{H_1} \cong 125\,\gev$ and $M_A = M_{H^\pm} \ge 180\,\gev$. In the
 lower-mass branch, $B1$, the plot of $M_{H_2}$ vs.~$M_A = M_{H^\pm}$ is
 reminiscent of the left panel in Fig.~\ref{fig:V1masses}. This behavior is
 not the result of a simple sum rule like Eq.~(\ref{eq:MHsum}), but the cause
 is much the same: requiring $M_{H_1} = 125\,\gev$ restricts $M_{H_2}$ to
 small values for large $M_A = M_{H^\pm}$. In this branch, which extends over
 $180 \le M_A = M_{H^\pm} \simle 380\,\gev$, $M_{H_2}$ starts near 550~\GeV,
 rises to 700~GeV and then drops rapidly to near zero at
 $M^*_A \cong 380\,\gev$. From there, branch~$B2$ rises rapidly and grows
 together indefinitely with the increasing input $M_{H'}$. For reasons we
 discuss in Sec.~IIc, we consider only branch $B1$ to be physically
 meaningful.

 As in perturbation theory in ordinary quantum mechanics, determining the
 mass eigenvalues to $\CO(V_2)$ requires that we know the eigenvectors $H_1$
 and $H_2$ only to $\CO(V_1)$. To that extent, what we have already stated
 about the degree of Higgs alignment --- that $H_1$ very nearly has SM
 couplings and that such processes as $H',A \to W^+W^-$, $ZZ$ and $HZ$ are
 greatly suppressed --- is still correct. Furthermore, alignment remains
 strong if we use the $\CO(V_1)$ approximation to just the numerator of
 Eq.~(\ref{eq:tandelta}) for $\tan 2\delta$ (see Eq.~(\ref{eq:del2})).

 In Sec.~III we follow a different approach to calculating the eigenvalues of
 $\CM^2_{0^+}$. It requires that the full two-loop effective potential,
 $V_0 + V_1 + V_2$, has a stable minimum. The program {\em
   Amoeba}~\cite{PresTeukVettFlan92} is used to find the regions of
 $V_{\rm eff}$ for which $\CM^2_{0^+}$ is positive-definite. We vary
 $M_A = M_{H^\pm}$ and $M_{H'}$ and require that $M_{H_1}$ or $M_{H_2}$ is
 equal $125\,\gev$. Only the solutions with the lighter eigenmass
 $M_{H_1} = 125\,\gev$ are consistent with LEP and LHC Higgs boson
 searches. We call this procedure the ``amoeba method''. As in Sec.~IIc,
 there are two regions of $M_A = M_{H^\pm}$ for this solution that we also
 call $B1$ and $B2$. Region~$B1$ extends from $M_A \cong 290\,\gev$ to
 $425\,\gev$ and $B2$ from $425\,\gev$ to about $600\,\gev$. Again, only
 region~$B1$ is physically meaningful. The behavior of the eigenvalue
 $M_{H_2}$ is quite similar in the $B1$ region of both methods as are the
 transitions between regions~$B1$ and~$B2$.
 
 Finally, in Sec.~IV we discuss the experimental implications of our two-loop
 studies, especially as they refer to the LHC experiments ATLAS and CMS.
 They are in good agreement with those in our previous
 papers~\cite{Lane:2018ycs},\cite{Lane:2019dbc},\cite{Eichten:2021qbm}: The
 BSM Higgs bosons are well within reach of the LHC today, but their discovery
 requires much improvement in the rejection of low-energy QCD backgrounds. We
 discuss two new search modes that have low rates, but also much lower
 backgrounds. Higgs alignment is respected with experimental violations and
 the corresponding suppression of many processes enjoyed by the SM Higgs
 below $\CO(1\%)$.

 \section*{II. The GW-2HDM at two-loops: the perturbative method} 

 Gildener and Weinberg's one-loop analysis~\cite{Gildener:1976ih} started
 from Eqs.~(\ref{eq:extrxa},\ref{eq:extrxb}). Because their analysis was
 intentionally model-independent, its main result was the very general, and
 very important, Eq.~(\ref{eq:M0sq1}) for the Higgs boson's mass. In the
 specific GW-2HDM, we can do more and extract other results. The most
 important ones so far are the sum rule~(\ref{eq:MHsum}) constraining the
 masses of the model's BSM Higgs bosons to be light and the degree to which
 Higgs alignment and the related suppression of BSM couplings to weak boson
 pairs and to a weak boson plus the SM Higgs~$H$. Determining the degrees to
 which they hold when extended to two loops motivate our present
 investigation.

We divide the discussion in this section into three parts: (a)~The formalism
for the extremal conditions and the two-loop contributions to the $C\!P$-even
scalar masses. This includes the generalization to two-loop order of
Eq.~(\ref{eq:lgw2}) relating the renormalization scale $\LGW$ to the
electroweak VEV~$v$. (b)~Calculations of $\LGW$ and $\CM^2_{0^+}$ in the
approximation of keeping {\em only} the all-scalar terms in
$V_2$. (c)~Determining the allowed ranges of the BSM masses,
$M_A = M_{H^\pm}$ and $M_{H_2}$ for $M_{H_1} = 125\,\gev$, and the degree
of Higgs alignment in the GW-2HDM. We refer to this procedure as the
``perturbative method'' because we discard terms that are formally of higher
order than two loops.

\subsection*{IIa. The two-loop formalism}

We extend the analysis in Ref.~\cite{Gildener:1976ih} to two-loop order
here. The key requirement of this is to retain only those terms that are at
most formally of second order in the loop expansion. The aligned basis,
Eq.~(\ref{eq:aligned}), is still the most suitable for this because, as we
shall see, the strictly two-loop corrections to Higgs alignment are small. In
the $C\!P$-conserving GW-2HDM, the only fields that can acquire a VEV are the
$C\!P$-even $H$ and $H'$. Therefore, through $\CO(V_2)$, and using
$\delta_1H = 0$, the extremal conditions are obtained from
\bea
\label{eq:extrVV2H}
\left.\frac{\partial(V_0 + V_1 + V_2)}{\partial
    H}\right\vert_{\langle\,\rangle + \delta_1H' + \delta_2H' + \delta_2H}
= 0,\\
\label{eq:extrVV2Hpr} \left.\frac{\partial(V_0 + V_1 + V_2)}{\partial H'}
\right\vert_{\langle\,\rangle + \delta_1H' + \delta_2H' + \delta_2H} = 0,
\eea
where, again, $\langle\,\rangle$ means the tree-level VEVs
$\langle H\rangle = v$ and $\langle H'\rangle = 0$. Using the
vanishing derivatives of $V_0$ in Eqs.~(\ref{eq:tree},\ref{eq:mevecH}) and
$(\partial^3 V_0/\partial H^3)_{\langle\,\rangle}  =
 (\partial^4 V_0/\partial H^4)_{\langle\,\rangle} = 0$, we get:

\bea
\label{eq:V2H}
0 &=& \left.\frac{\partial(V_0 + V_1 + V_2)}{\partial
    H}\right\vert_{\langle\,\rangle + \delta_1H' + \delta_2H' + \delta_2H}
\nn\\\nn\\
&=&  \left.\frac{\partial V_1}{\partial H}\right\vert_{\langle\,\rangle}
    +\half\left.\frac{\partial^3 V_0}{\partial H\partial
    H^{\prime\,2}}\right\vert_{\langle\,\rangle}(\delta_1H')^2
+ \left.\frac{\partial^2 V_1}{\partial H\partial H'}
\right\vert_{\langle\,\rangle} \delta_1H'
+ \left.\frac{\partial V_2}{\partial
    H}\right\vert_{\langle\,\rangle}\nn\\\nn\\
&=& \frac{1}{16\pi^2 v}\sum_n\alpha_n M_n^4\left(\ln\frac{M_n^2}{\LGW^2} +\half
  - k_n\right) \nn\\
&& - \frac{\alpha_t M_t^4 \tan\beta\,\delta_1H'}{8\pi^2 v^2}
 \left(\ln\frac{M_t^2}{\LGW^2} +\frac{3}{2} - k_t\right)
 + \left.\frac{\partial V_2}{\partial H}\right\vert_{\langle\,\rangle};
%
%
 \eea
 \bea
\label{eq:V2Hpr}
0 &=& \left.\frac{\partial(V_0 + V_1 + V_2)}{\partial
    H'}\right\vert_{\langle\,\rangle + \delta_1H' + \delta_2H' +
  \delta_2H}\nn\\ \nn\\
&=&
\left.\frac{\partial^2 V_0}{\partial
  H^{\prime\,2}}\right\vert_{\langle\,\rangle}\delta_1H'
+ \left.\frac{\partial V_1}{\partial H'}\right\vert_{\langle\,\rangle}
+ \left.\frac{\partial^2 V_0}{\partial
H^{\prime\,2}}\right\vert_{\langle\,\rangle}\delta_2H'
+ \left.\frac{\partial^2 V_1}{\partial H^{\prime\,2}}
\right\vert_{\langle\,\rangle} \delta_1H' 
 + \half\left.\frac{\partial^3 V_0}{\partial
        H^{\prime\,3}}\right\vert_{\langle\,\rangle}\,(\delta_1H')^2
+ \left.\frac{\partial V_2}{\partial
    H'}\right\vert_{\langle\,\rangle}\nn\\\nn\\
&=& \biggl[\frac{\alpha_t M_t^4(1+3\tan^2\beta)}
{32\pi^2 v^2} \left(\ln\frac{M_t^2}{\LGW^2} +\frac{1}{2} - k_t\right)
    +\frac{\alpha_t M_t^4 \tan^2\beta}{8\pi^2 v^2}\nn\\
&&  +\frac{1}{16\pi^2 v}\sum_n\alpha_n M_n^4\left(\ln\frac{M_n^2}{\LGW^2}
    +\half - k_n\right)\biggr]\delta_1H' + M^2_{H'}\delta_2H'
+\left.\frac{\partial V_2}{\partial
    H'}\right\vert_{\langle\,\rangle}.\hspace{1.0cm}
\eea
Here, we used Eq.~(\ref{eq:Vzalign}) to calculate the derivatives of $V_0$,
the definition of the $\CO(V_1)$ shift $\delta_1H'$ in the VEV of $H'$, in
Eq.~(\ref{eq:extrx2}), and the following:
\bea\label{eq:derivs1}
&& \left.\frac{\partial V_1}{\partial H}\right\vert_{\langle\,\rangle}
 = \frac{1}{16\pi^2 v}\sum_n\alpha_n M_n^4\left(\ln\frac{M_n^2}{\LGW^2} +\half
   - k_n\right)\\
%
%
\label{eq:derivs2}
&& \left.\frac{\partial^2 V_1}{\partial H\partial
    H'}\right\vert_{\langle\,\rangle} =
 -\frac{3\alpha_t M_t^4 \tan\beta}{16\pi^2 v^2}\left(\ln\frac{M_t^2}{\LGW^2}
     +\frac{7}{6} - k_t\right),\\
\label{eq:derivs3}
&& \left.\frac{\partial^2 V_1}{\partial H^{\prime\,2}}
\right\vert_{\langle\,\rangle} = \frac{\alpha_t
  M_t^4\left(3\tan^2\beta - 1\right)}{16\pi^2 v^2}
   \left(\ln\frac{M_t^2}{\LGW^2} +\half - k_t\right)
   +\frac{2\alpha_t M_t^4\tan^2\beta}{16\pi^2 v^2}\nn\\
&&\hspace{2.0cm} + \frac{1}{16\pi^2 v^2}\sum_n\alpha_n M_n^4
\left(\ln\frac{M_n^2}{\LGW^2} +\half - k_n\right).
\eea
Every term on the right side of Eqs.~(\ref{eq:V2H},\ref{eq:V2Hpr}) is of
$\CO(V_2)$ or, sometimes more explicitly, $\CO(\kappa^2)$, where
\be\label{eq:kapdef}
\kappa = \frac{1}{16\pi^2}.
\ee
This is because, as stated below Eq.~(\ref{eq:lamconds}), the extremal
conditions in each order of the loop expansion of $V_{\rm eff}$ are enforced
in all orders of the loop expansion~\cite{Gildener:1976ih}. That means,
e.g., that the right side of Eq.~(\ref{eq:derivs1} and the third term in
Eq.~(\ref{eq:derivs3} are $\CO(V_2)$. This will provide an $\CO(V_2)$
correction to $\LGW$.



The dominant $\CO(\kappa^2)$ corrections to the extremal conditions will come
from the derivatives of $V_2$ itself with respect to $H$ and $H'$.
Equation~(\ref{eq:V2H}) determines the $\CO(V_1) = \CO(\kappa)$ correction to
$\LGW$. From now on, we denote the renormalization scale by $\LGW$ {\em only}
in terms that are otherwise of $\CO(V_1)$. In those terms, the $\CO(V_1)$
part of $\LGW$ will produce an $\CO(\kappa^2)$ contribution. In terms that
are already $\CO(V_2)$, we use the $\CO(\kappa^0)$ scale
$\Lambda_0 = v\exp({\thalf}(A/B + {\thalf}))$ from Eq.~(\ref{eq:lgw2}). We
obtain the following expression for $\LGW$ (in which we still use
$M^2_H = (\partial^2 V_1/\partial H^2)_{\langle\,\rangle} = \sum_n \alpha_n
M_n^4/8\pi^2v^2$):
\bea\label{eq:LGW}
\LGW &=& \Lambda_0\exp\left\{\frac{2}{M^2_H v}
  \left[\frac{\alpha_t M^4_t\tan\beta\,\delta_1H'}{8\pi^2 v^2}
  \left(\log\frac{M^2_t}{\Lambda^2_0} +\frac{3}{2} - k_t\right)
  - \left.\frac{\partial V_2}{\partial H}
  \right\vert_{\langle\,\rangle}\right]\right\} \nn\\
&\cong& \Lambda_0\left[1 + \frac{\alpha_t M^4_t \tan\beta\,\delta_1H'}
  {4\pi^2 v^3 M^2_H}\left(\log\frac{M^2_t}{\Lambda^2_0} +\frac{3}{2} -
    k_t\right) - \frac{2}{M^2_H v}\left.\frac{\partial V_2}{\partial H}
      \right\vert_{\langle\,\rangle}\right].\hspace{1.00cm}
\eea
This correction to $\Lambda_0$ is $\CO(\kappa)$ because $M^2_H =
\CO(\kappa)$.

Equation~(\ref{eq:V2Hpr}) determines the $\CO(\kappa^2)$ contribution
$\delta_2H'$ to $\delta H'$:
\bea\label{eq:del2Hpr}
\delta_2H' &=& -\frac{1}{M_{H'}^2}\left[\frac{1}{2}\left.\frac{\partial^3
      V_0}{\partial H^{\prime\,3}}\right\vert_{\langle\,\rangle}(\delta_1H')^2
+\left.\frac{\partial^2 V_1}{\partial
    H^{\prime\,2}}\right\vert_{\langle\,\rangle} \delta_1H'
+\left.\frac{\partial V_2}{\partial
    H'}\right\vert_{\langle\,\rangle}\right]\nn\\
=&& \hspace{-0.75cm} -\frac{1}{M_{H'}^2}\left[\left(\frac{\alpha_t M_t^4
      (1+3\tan^2\beta)}{32\pi^2 v^2}\left(\ln\frac{M_t^2}{\Lambda_0^2}
      + \frac{1}{2} - k_t\right) + \frac{\alpha_t M_t^4\tan^2\beta}{8\pi^2
      v^2}\right)\delta_1H' + \left.\frac{\partial
    V_2}{\partial H'}\right\vert_{\langle\,\rangle}\right].\hspace{0.75cm}
\eea

The shift $\delta_2H$ does not appear in Eqs.~(\ref{eq:V2H},\ref{eq:V2Hpr}).
It could do so to $\CO(V_2)$ only by multiplying
$(\partial^2 V_0/\partial H^2)_{\langle\,\rangle} = 0$ and
$(\partial^2V_0/\partial H\partial H')_{\langle\,\rangle} = 0$ by
$\delta_2H$. Since it is undetermined, we use it to keep $v$ fixed. That is,
we require\footnote{Eq.~(\ref{eq:vdef}) is correct through $\CO(V_2)$.}
\be\label{eq:vdef}
v^2 = (v + \delta_2H)^2 + (\delta_1H' + \delta_2H')^2 \Longrightarrow
\delta_2H = -(\delta_1H')^2/2v.
\ee
%


Now turn to the elements of the $C\!P$-even squared mass matrix in
$\CO(V_2)$. With an obvious notation, they are:
\bea\label{eq:MV2}
\left(\CM^2_{H_iH_j}\right)_2 &=& \left.\frac{\partial^2(V_0 + V_1 + V_2)}
  {\partial H_i \partial H_j}\right\vert_{\langle\,\rangle + \delta H + \delta
  H'}\nn\\\nn\\
&=& \left.\frac{\partial^2 V_0}{\partial H_i \partial H_j}
    \right\vert_{\langle\,\rangle} +
    \left.\frac{\partial^3 V_0}{\partial H_i\partial H_j\partial H_k}
    \right\vert_{\langle\,\rangle}(\delta_1H_k + \delta_2H_k) +
    \half\left.\frac{\partial^4 V_0}{\partial H_i\partial H_j\partial
    H_k\partial H_l}\right\vert_{\langle\,\rangle}\delta_1H_k\,\delta_1H_l\nn\\
&+&  \left.\frac{\partial^2 V_1}{\partial H_i \partial H_j}
    \right\vert_{\langle\,\rangle} +
     \left.\frac{\partial^3 V_1}{\partial H_i\partial H_j\partial H_k}
    \right\vert_{\langle\,\rangle}\delta_1H_k +
    \left.\frac{\partial^2 V_2}{\partial H_i \partial H_j}
    \right\vert_{\langle\,\rangle}.
\eea
Then:
\bea
\label{eq:HH}
\left(\CM^2_{HH}\right)_2 &=&
  M^2_H + M^2_{H'}\left(\frac{\delta_1H'}{v}\right)^2
  -\frac{3\alpha_t M_t^4 \,\tan\beta}{8\pi^2 v^2}
  \left(\ln\frac{M_t^2}{\Lambda_0^2} + \frac{13}{6} - k_t\right)
  \frac{\delta_1H'}{v} +
  \left.\frac{\partial^2 V_2}{\partial H^2}\right\vert_{\langle\,\rangle}\nn\\
&=& M^2_H -\frac{5\alpha_t M_t^4\tan\beta}{16\pi^2 v^2}
\left(\ln\frac{M_t^2}{\Lambda_0^2} +\frac{5}{2} - k_t\right)\frac{\delta_1H'}{v} +
\left.\frac{\partial^2 V_2}{\partial H^2}\right\vert_{\langle\,\rangle};\\\nn\\
\label{eq:HHp}
\left(\CM^2_{HH'}\right)_2 &=& -\frac{\alpha_t M_t^4\tan\beta}{16\pi^2 v^2}
\left(\ln\frac{M_t^2}{\LGW^2} + \frac{5}{2} - k_t\right)\nn\\
&+& \left[M^2_H + \frac{3\alpha_t M_t^4(\tan^2\beta -1)}{32\pi^2
    v^2}\left(\ln\frac{M_t^2}{\Lambda_0^2}+\frac{1}{2}-k_t\right)
+ \frac{\alpha_t M_t^4(3\tan^2\beta-1)}{8\pi^2
  v^2}\right]\frac{\delta_1H'}{v} \nn\\
&-&\frac{2}{v}\left.\frac{\partial V_2}{\partial
    H'}\right\vert_{\langle\,\rangle}  + \left.\frac{\partial^2 V_2}{\partial H
   \partial H'}\right\vert_{\langle\,\rangle};
\\\nn\\
\label{eq:HpHp}
\left(\CM^2_{H'H'}\right)_2 &=& M_{H'}^2 + \frac{\alpha_t M_t^4}{8\pi^2 v^2}
        \left(\ln\frac{M_t^2}{\LGW^2}+\frac{1}{2}-k_t+\tan^2\beta\right) \nn\\
&-& \left[\frac{7\alpha_t M_t^4\tan\beta}{16\pi^2 v^2}
  \left(\ln\frac{M_t^2}{\Lambda_0^2}+\frac{1}{2}-k_t\right)
+ \frac{\alpha_tM_t^4\tan\beta(3 + 2\tan^2\beta)}{8\pi^2 v^2}\right]
\frac{\delta_1H'}{v} \nn\\
&-&\frac{6\cot 2\beta}{v}\left.\frac{\partial V_2}{\partial
    H'}\right\vert_{\langle\,\rangle}  + \left.\frac{\partial^2 V_2}{\partial
  \Hpt}\right\vert_{\langle\,\rangle}.
\eea
Equation~(\ref{eq:vdef}) for $\delta_2H$ was used in calculating
$\CM^2_{H'H'}$.

When determining the eigenmasses $M^2_{H_1,H_2}$ in Eqs.~(\ref{eq:H1H2evals}) to
$\CO(V_2)$, {\em only} the $\CO(\kappa)$ term in $\CM^2_{HH'}$ should be kept
(using $\Lambda_0$) and then multiplied by
$\sin\delta_1\cos\delta_1 = \CO(\kappa)$. For the same reason, the term
$\sin^2\delta_1 \CM^2_{HH}$ in $M^2_{H_2}$ should be dropped. The
eigenvalues of $\CM^2_{0^+}$ to $\CO(V_2)$ are then
\bea\label{eq:H1H2evalsV2}
M^2_{H_1} &=&
\left(\CM^2_{HH}\right)_2\cos^2\delta_1 +
\left(\CM^2_{H'H'}\right)_0\sin^2\delta_1 -
 2\left(\CM^2_{HH'}\right)_1\sin\delta_1 \cos\delta_1,\nn\\
M^2_{H_2} &=&
\left(\CM^2_{H'H'}\right)_2\cos^2\delta_1 +
  2(\CM^2_{HH'})_1\sin\delta_1\cos\delta_1,
\eea
where only the $\CO(V_1)$ part of
$\tan 2\delta_1 \cong 2\CM^2_{HH'}/\CM^2_{H'H'}$ in Eq.~(\ref{eq:tandelta})is
used. The left side of Fig.~\ref{fig:V1masses} shows that these are good
approximations.

On the other hand, for the purpose of determining the eigenvectors $H_1$ and
$H_2$ and their degree of alignment from the $\CO(V_2)$ version of
Eq.~(\ref{eq:H1H2evecs}), we use $\delta_2$ defined by
\be\label{eq:del2}
\tan 2\delta_2 = \frac{\left(2\CM^2_{HH'}\right)_1}
{\left(\CM^2_{H'H'}-\CM^2_{HH}\right)_2}
\ee
because this approximation is numerically closer to $\delta_1$ than that
which results from expanding $\tan 2\delta$ to $\CO(\kappa^2)$.

\subsection*{IIb. The scalar approximation}

There are five general types of contributions to the two-loop potential $V_2$
for the GW-2HDM and similar electroweak models; see Secs.~2 and~4 of
Ref.~\cite{Martin:2001vx} for details of the interactions and the two-loop
integrals.
\begin{itemize}

\item[1.)] Scalar graphs consisting of ``cracked-egg'' two-vertex graphs with
  three scalars emanating from one interaction vertex and propagating to the
  other (SSS), and ``figure-eight'' graphs with two separate one-loop graphs
  (each loop as in $V_1$) stuck together at a single vertex with the
  appropriate quartic coupling (SS). These contributions arise from the
  scalar potential $V_0$ in Eq.~(\ref{eq:Vzalign}), as described below.

\item[2.)] Cracked-egg fermion loops, induced by Yukawa interactions, with a
  scalar exchanged between the two vertices (FFS); only the top and bottom
  quarks contribute significantly to the loop integrals.

\item[3.)] From the electroweak gauge interactions of the scalars there are
  cracked-egg scalar loops with an electroweak gauge boson exchanged between
  the two vertices (SSV) and figure-eight graphs with a scalar loop and a
  gauge loop (VS). There are also cracked-egg electroweak gauge loops with a
  scalar exchanged between the vertices (VVS).

\item[4.)] Cracked-egg fermion loops with an electroweak boson or QCD gluon
  exchanged between the two vertices (FFV); again, only $t$ and $b$ quarks
  contribute substantially.

\item[5.)] Pure gauge-boson (including ghosts) cracked-egg and figure-eight
  loops (gauge).

\end{itemize}
\begin{figure}[h!]
  \begin{center}
    \includegraphics[width = 5.00in,height=3.00in]{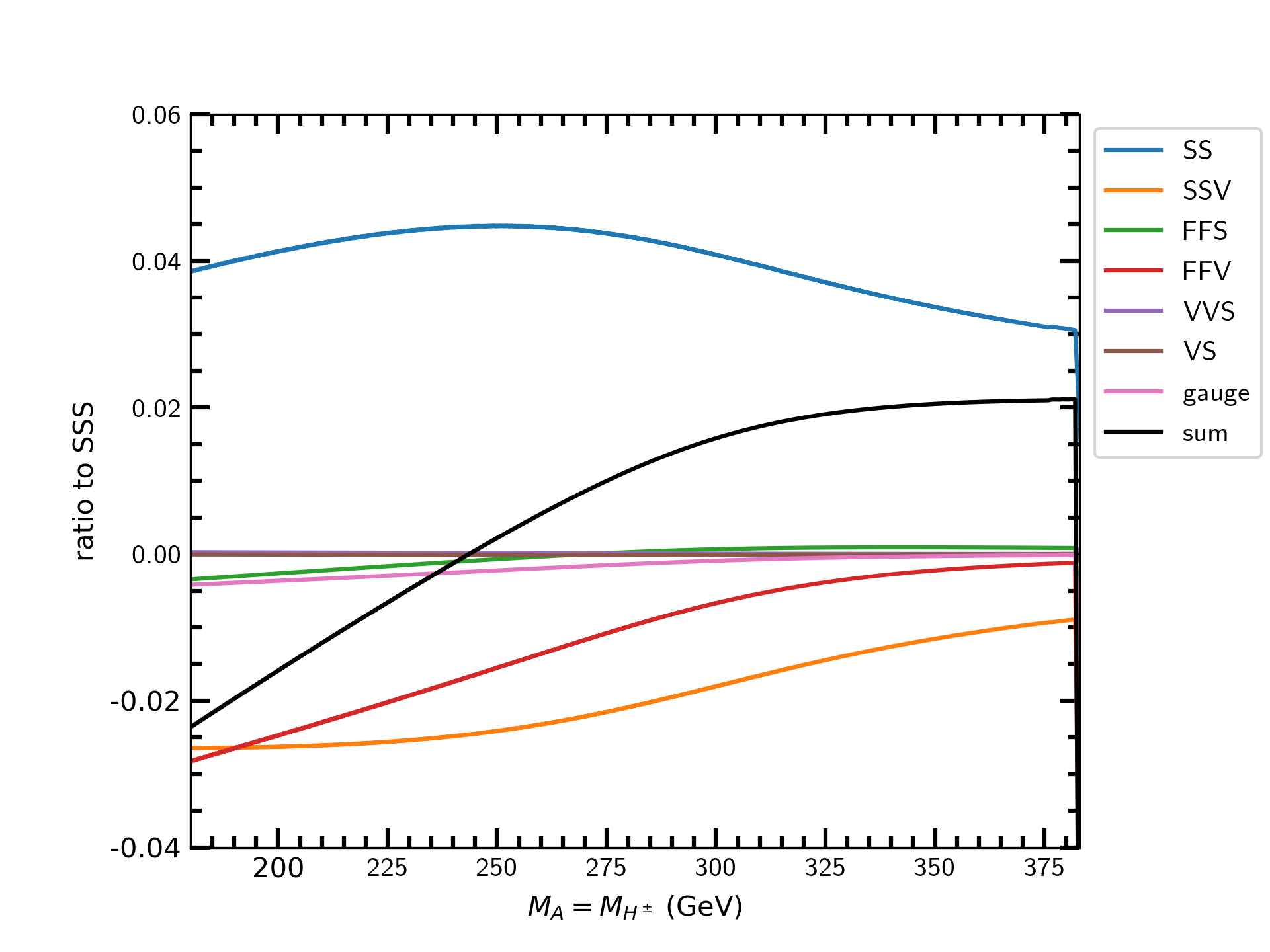}
    \caption{Ratios to the SSS cracked-egg contribution of the SS
      figure-eight, SSV, FFS, FFV, VVS, VS figure-eight, and gauge
      contributions to $V_2$ for
      $180\,\gev\le M_A = M_{H^\pm} \simle 380\,\gev$, the
      region of branch $B1$ in Fig.~\ref{fig:MH2}. The black curve is the sum
    of the eight ratios. In addition to $V_{SSS}$ and $V_{SS}$, these
    two-loop potentials are taken from Martin~\cite{Martin:2001vx}.}
      \label{fig:B1ratios}
    \end{center}
  \end{figure}
  Of these five types of contributions to $V_2$, the scalar (SSS and SS)
  graphs are by far the most important because the BSM Higgs masses set their
  magnitudes.\footnote{As in $V_1$, the tree-level masses are used for all
    the scalars, gauge bosons and fermions propagating in these
    loops.}$^{,}$\footnote{The cracked-egg scalar graphs are much larger than
    the figure eights; see also Ref.~\cite{Braathen:2020vwo}.} Therefore, we
  approximate $V_2$ by its scalar contributions. This approximation is good
  to about ${2\%}$ over the entire range of branch $B1$; see
  Fig.~\ref{fig:B1ratios}.

  The SSS couplings descend from the quartic couplings in $V_0$ of
  Eq.~(\ref{eq:Vzalign}) by shifting the scalar quantum fields by their
  classical counterparts~\cite{Jackiw:1974cv}.\footnote{The only other
    cracked-egg graphs with field-dependent couplings are VVS with V an
    electroweak boson. They descend from the quartic electroweak interactions
    and are of order a squared electroweak coupling times $H_c$ or $H'_c$.}
  Following Ref.~\cite{Martin:2001vx}, it is convenient to use real scalar
  (and electroweak boson) fields for this discussion:
\bea\label{eq:Ri}
R_1 &=& H',\,\, R_2 = A,\,\, R_3 = h_1,\,\, R_4 = h_2;\nn\\
R_5 &=& H,\,\, R_6 = z,\,\, R_7 = w_1,\,\, R_8 = w_2.\nn\\
A^1_\mu &=& W^1_\mu,\,\,A^2_\mu = W^2_\mu,\,\, A^3_\mu = Z_\mu,\,\,A^4_\mu =
A_\mu\,\,{\text{(the photon)}}.
\eea
Here, $H^\pm =(h_1 \pm i h_2)/\sqrt{2}$, $w^\pm = (w_1 \pm i w_2)/\sqrt{2}$
and $W^\pm_\mu = (A^1_\mu \pm i A^2_\mu)/\sqrt{2}$. Because our interest in
calculating $V_2$ is to see its effect on $M_{H_2}$ as we vary
$M_A = M_{H^\pm}$ (as the sum rule~(\ref{eq:MHsum}) did in $\CO(V_1)$) and on
the $H$--$H'$ mixing determining the departure from Higgs alignment, we shift
only the two scalar fields that can get a $C\!P$-conserving VEV, $H$ and
$H'$:\footnote{Strictly speaking, the derivatives with respect to $H$ and
  $H'$ in Secs.~I and IIa were with respect to $H_c$ and $H'_c $, but the
  results there do not depend on this point; also see footnote~9.}
\be\label{eq:Hshifts}
R_5 = H \to R_5 + H_c, \quad R_1 = H' \to R_1 + H'_c.
\ee
The cubic-scalar interactions are those that are first order in
$H_c$~or~$H'_c$.\footnote{The terms quadratic in the classical fields gave
  rise to the one-loop potential $V_1$~\cite{Jackiw:1974cv}.} We indicate
these couplings with an {\em overbar}, $\blam_{ijk}$, as we did for the
field-dependent masses $\bM_n^2$ in Eq.~(\ref{eq:bkgM}).

The scalar interactions used in constructing $V_2$ of the GW-2HDM are then
\be
\label{eq:VS}
V_S = \tsixth\blam_{ijk} R_iR_jR_k + \ttwofour\lambda_{ijkl}R_iR_jR_kR_l,
\ee
where repeated indices are summed over and the prefactors of
${\textstyle{\frac{1}{6}}}$ and ${\textstyle{\frac{1}{24}}}$ are choices of
convenience made in Ref.~\cite{Martin:2001vx}. The triple-scalar
couplings consistent with these normalizations are:
\bea\label{eq:blamijk}
\blam_{111} &=& 6M^2_{H'}\cot 2\beta(H_c + 2H'_c\cot2\beta)/v^2;\nn\\ \nn\\
\blam_{122} &=& \blam_{133} = \blam_{144} =  2M^2_{H'}\cot 2\beta(H_c +
   2H'_c\cot2\beta)/v^2;\nn\\ \nn\\
\blam_{115}&=& 2M^2_{H'}(H_c + 3H'_c \cot 2\beta)/v^2,\,\,\,
\blam_{225} = 2(M^2_{H'} H'_c\cot 2\beta + M^2_A H_c)/v^2;\nn\\ \nn\\
\blam_{335} &=& \blam_{445} = 2(M^2_{H'} H'_c\cot 2\beta + M^2_{H^\pm}H_c)/v^2;
\nn\\ \nn\\
\blam_{155} &=& 2M^2_{H'}H'_c/v^2,\,\,\,
\blam_{166} = 2M^2_A H'_c/v^2,\,\,\, \blam_{177} = \blam_{188} =
2M^2_{H^\pm}H'_c/v^2;\nn\\ \nn\\
\blam_{126} &=& (M^2_{H'} - M^2_A)H_c/v^2,\,\,\,
\blam_{256} = (M^2_{H'} - M^2_A)H'_c/v^2;\nn\\ \nn\\
\blam_{137} &=& \blam_{148} =((M^2_{H'} - M^2_{H^\pm})H_c + 2M^2_{H'}H'_c\cot
    2\beta)/v^2; \nn\\ \nn\\
\blam_{357} &=& \blam_{458} = (M^2_{H'} - M^2_{H^\pm})H'_c/v^2;\nn\\ \nn\\
\blam_{238} &=& -\blam_{247} = (M^2_A - M^2_{H^\pm})H_c/v^2,\,\,\,
\blam_{467} = -\blam_{368} = (M^2_A - M^2_{H^\pm})H'_c/v^2.\hspace{1.0cm}
\eea
In Eq.~(\ref{eq:VS}), $\blam_{ijk}$ appears six times, $\blam_{iij}$ three
times, and $\blam_{iii}$ once.

The $\lambda_{ijkl}$ in Eq.~(\ref{eq:VS}) are the quartic scalar couplings in
Eq.~(\ref{eq:Vzalign}). Because of the figure-eight structure of the two-loop
graphs to which they contribute ($V^{(2)}_{SS}$ in Eq.~(\ref{eq:SS}) below),
only terms with $\lambda_{iiii}$ and $\lambda_{iijj}$ with $i\neq j$ in $V_S$
are used; $\lambda_{iiii}$ contributes to one term in $V_S$ and
$\lambda_{iijj}$ (with $i< j$, e.g.)  contributes to six terms there. They
are:

\bea\label{eq:lamijkl}
%
%
\lambda_{1122} &=& \lambda_{1133} = \lambda_{1144} = 4M_{H'}^2\cot^2
  2\beta/v^2;\nn\\
\lambda_{2233} &=& \lambda_{2244} = \lambda_{3344} = 4M_{H'}^2\cot^2
  2\beta/v^2;\nn\\
\lambda_{1111} &=& \lambda_{2222} = \lambda_{3333} = \lambda_{4444} =
12 M_{H'}^2\cot^2 2\beta/v^2.
\eea

The two-loop effective potential in the scalar approximation is given
by~\cite{Martin:2001vx}
\bea\label{eq:SPMV2}
V_2 &=& \kappa^2 \left(V^{(2)}_{SSS} + V^{(2)}_{SS}\right),
\eea
where, in terms of the couplings and field-dependent masses specified above:
\bea
\label{eq:SSS}
 V^{(2)}_{SSS} &=& -\ttwelfth \blam^2_{ijk}\, I(\bM^2_i,\bM^2_j,\bM^2_k),
   \\
\label{eq:SS}
 V^{(2)}_{SS} &=& \teighth \lambda_{iijj}\, J(\bM^2_i,\bM^2_j).
 \eea
 All indices on the right are summed over. The loop-integral functions\\
 $I(\bM^2_i,\bM^2_j,\bM^2_k)$ and $J(\bM^2_i,\bM^2_j)$ are defined in
 Ref.~\cite{Martin:2001vx}. They are symmetric under the interchange of their
 arguments. Therefore, there are two equal terms in $V^{(2)}_{SS}$ with
 $\lambda_{iijj}$ with $i \neq j$. In using Martin's formulas for the
 $I$-integral, and various massless limits of it, it is important to note
 that the arguments are ordered as $M_i^2 \le M_j^2 \le M_k^2$. Martin
 included in the definition of these functions all factors associated with
 the evaluation of Feynman diagrams, including fermion-loop minus
 signs.\footnote{Only particles that become massive at tree level contribute
   to the figure-eight loop function $J(\bM^2_i,\bM^2_j)$ so that, e.g., we
   omitted $\lambda_{1155}$ in Eqs.~(\ref{eq:lamijkl}).}

\subsection*{IIc. Numerical results for $H',A,H^\pm$ masses}

A glance at Eqs.~(\ref{eq:HH},\ref{eq:HHp},\ref{eq:HpHp}) for the $C\!P$-even
masses will convince the reader that a simple, useful generalization to
$\CO(V_2)$ of the sum rule~(\ref{eq:MHsum}) is out of the question. This is
so even in the approximation of keeping only the all-scalar graphs. To study
the mass $M_{H_2}$ as a function of the other scalar masses,
$M_A = M_{H^\pm}$ (see footnote~2), $M_{H'}$ and $M_{H_1}$, we use the
following algorithm:
\begin{itemize}
\item[1.)] Increment $M_A = M_{H^\pm}$ from $180\,\gev$ to
  $1\,\tev$.\footnote{These are the approximate lower bound set by searches
    for $H^\pm \to \tau^\pm \nu_\tau$, $c \bar b$ and $c \bar s$ at LEP and
    the LHC and well above the upper bound of $\simeq 500\,\gev$ expected
    from the one-loop sum rule.} We use
  $\tan\beta = 0.50$~\cite{Lane:2018ycs}.
\item[2.)] For each value of $M_A$, increment $M_{H'}$ from $10\,\gev$ to
  $1\,\tev$.
\item[3.)] Calculate the $\CO(\kappa)$ renormalization scale $\LGW$
  (Eq.~(\ref{eq:LGW})) and the two-loop shifts in the VEVs of $H$ and $H'$.
\item[4.)] Calculate the $\CO(V_2)$ elements $(\CM^2_{H_iH_j})_2$
  consistent with the extremal conditions solved for $\LGW$ and $\delta_2H'$.
  Then diagonalize them to $\CO(\kappa^2)$ using Eq.~(\ref{eq:H1H2evalsV2})
  with $\delta_1$ given by the $\CO(V_1)$ approximation to
  Eq.~(\ref{eq:tandelta}). For comparison, we also calculated the eigenvalues
  and eigenvectors of $(\CM^2_{H_iH_j})_2$ using the approximate
  two-loop $H$--$H'$ mixing angle $\delta_2$ in Eq.~(\ref{eq:del2}). This
  made no discernible difference in the masses and the degree of Higgs
  alignment.
\item[5.)]  We select for plotting the $C\!P$-even eigenmasses satisfying:

  {(a)} The one-loop Higgs mass-squared
  $M^2_H = \sum_n\alpha_n M^4_n/8\pi^2v^2 > 0.$

  {(b)} $\left(\CM^2_{H'H'}\right)_2 > 0.$

  {(c)} $|M^2_{H_1} - (125\,\gev)^2| \le 1250\,\gev^2$.

\end{itemize}

These conditions always yield positive $M^2_{H_1,H_2}$. For fixed $M_A$, the
selections are usually multi-valued, satisfied for several values of
$M_{H'}$. We plot the selection having $M_{H_1}$ closest to $125\,\gev$. We
also plot the renormalization scales $\Lambda_0$ and $\LGW$. As noted
earlier, this procedure does not guarantee that $\det(\CM^2_{0^+}) > 0$. In
this analysis, $\det(\CM^2_{0^+}) > 0$ only for $M_A < 260\,\gev$.

\begin{figure}[ht!]
 \begin{center}
\includegraphics[width=2.65in,
height=2.65in]{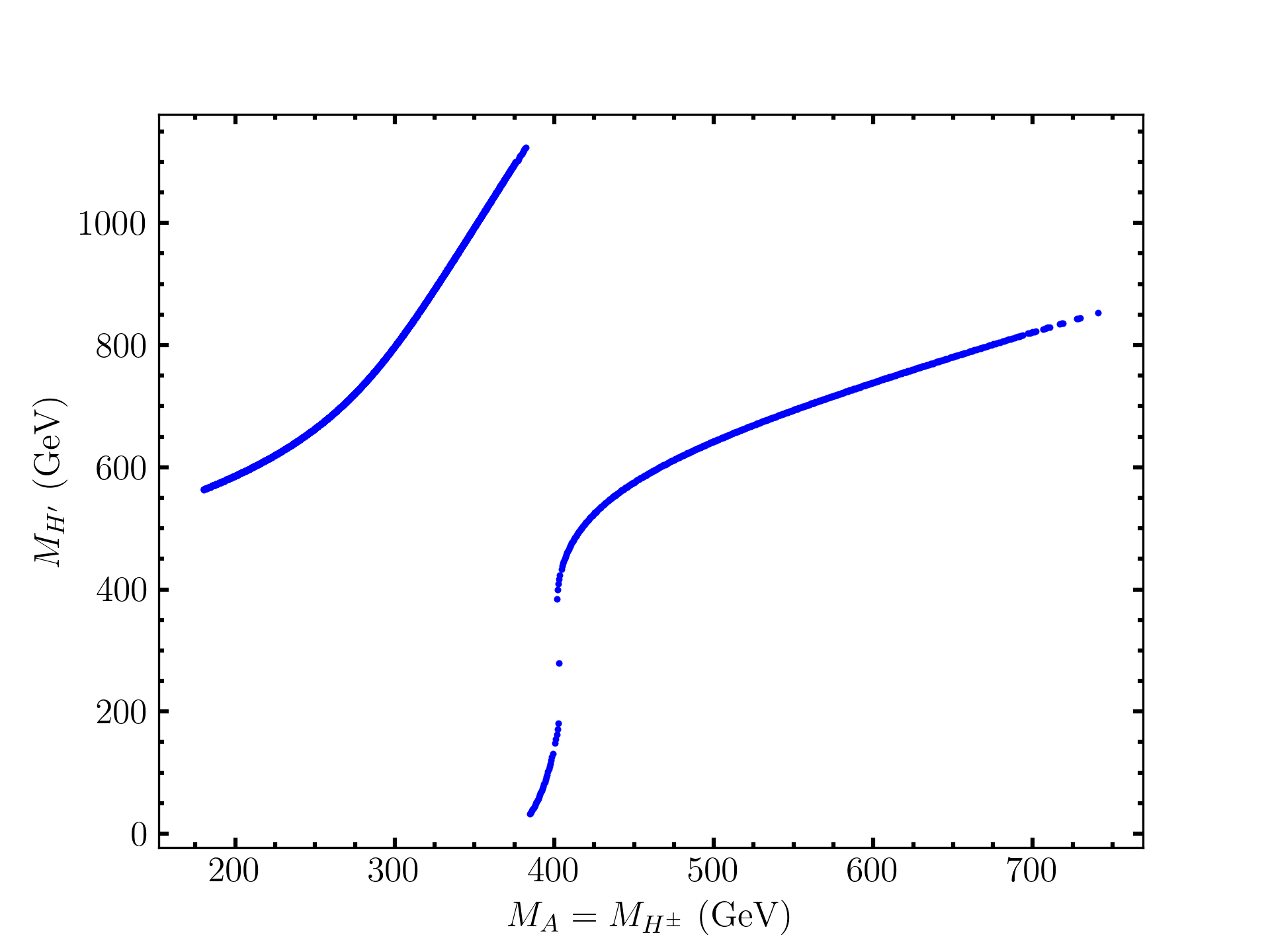}
\includegraphics[width=2.65in,
height=2.65in]{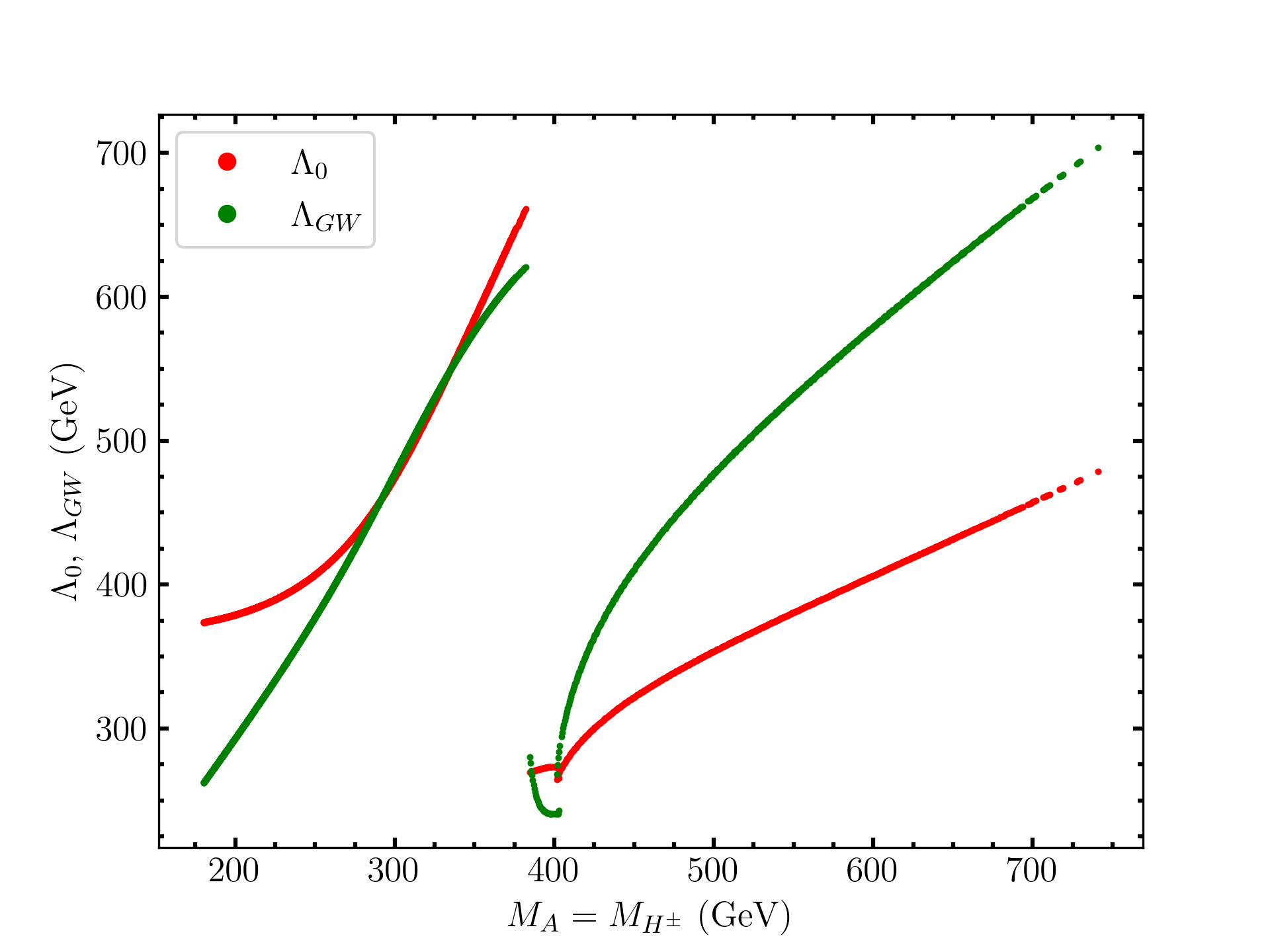}
\caption{Left: The BSM Higgs masses $M_{H'}$ vs. $M_A = M_H^\pm$ with the
  two-loop Higgs boson mass $M_{H_1}$ fixed near $125\,\gev$ as described in
  the text. Here and below, $\tan\beta = 0.50$.
  Right: The renormalization scale
  $\Lambda_0$ (red) calculated to zero-loop order from
  Eqs.~(\ref{eq:lgw2},\ref{eq:lgw3}) and the one-loop scale $\LGW$ (green)
  from Eq.~(\ref{eq:LGW}). The two branches, $B1$ and $B2$, of $M_{H'}$ and
  of the renormalization scales are discussed in the text. The transition
  between them occurs at $M_A^*({\rm pert.}) \cong 380\,\gev$.}
  \label{fig:MHpr_LGW}
 \end{center}
\end{figure}

Figure~\ref{fig:MHpr_LGW} shows $M_{H'}$ vs.~$M_A$ on the left and the
renormalization scales $\Lambda_0$ and $\LGW$ on the right. There are two
branches of each, $B1$ for $M_A < M_A^*({\rm pert.}) \cong 380\,\gev$ and
$B2$ for $M_A > M_A^*({\rm pert.})$. On the left, the values of $M_{H'}$ for
which $M_{H_1} \cong 125\,\gev$ in branch $B1$ start near $550\,\gev$ and
then rise approximately linearly with $M_A$ to over $1\,\tev$. This branch
ends abruptly at $M_A^*({\rm pert.})$. Branch $B2$ begins there near
$M_{H'} = 0$, rising quickly and then growing linearly with $M_A$ up to
$M_A \simeq 750\,\gev$ and $M_{H'} \simeq 825\,\gev$ where the data becomes
sparse because the algorithm conditions can no longer be satisfied. In branch
$B1$ of the right panel, the $\CO(\kappa)$ scale $\LGW$ starts below
$\Lambda_0$ and grows linearly with $M_A$, becoming almost equal to
$\Lambda_0$ at $M_A \cong 260\,\gev$ for the remainder of $B1$. In branch
$B2$, both scales grow linearly with $M_A$ over the range calculated, but
with a greater slope for $\LGW$.

\begin{figure}[ht!]
 \begin{center}
\includegraphics[width=2.65in,height=2.65in]{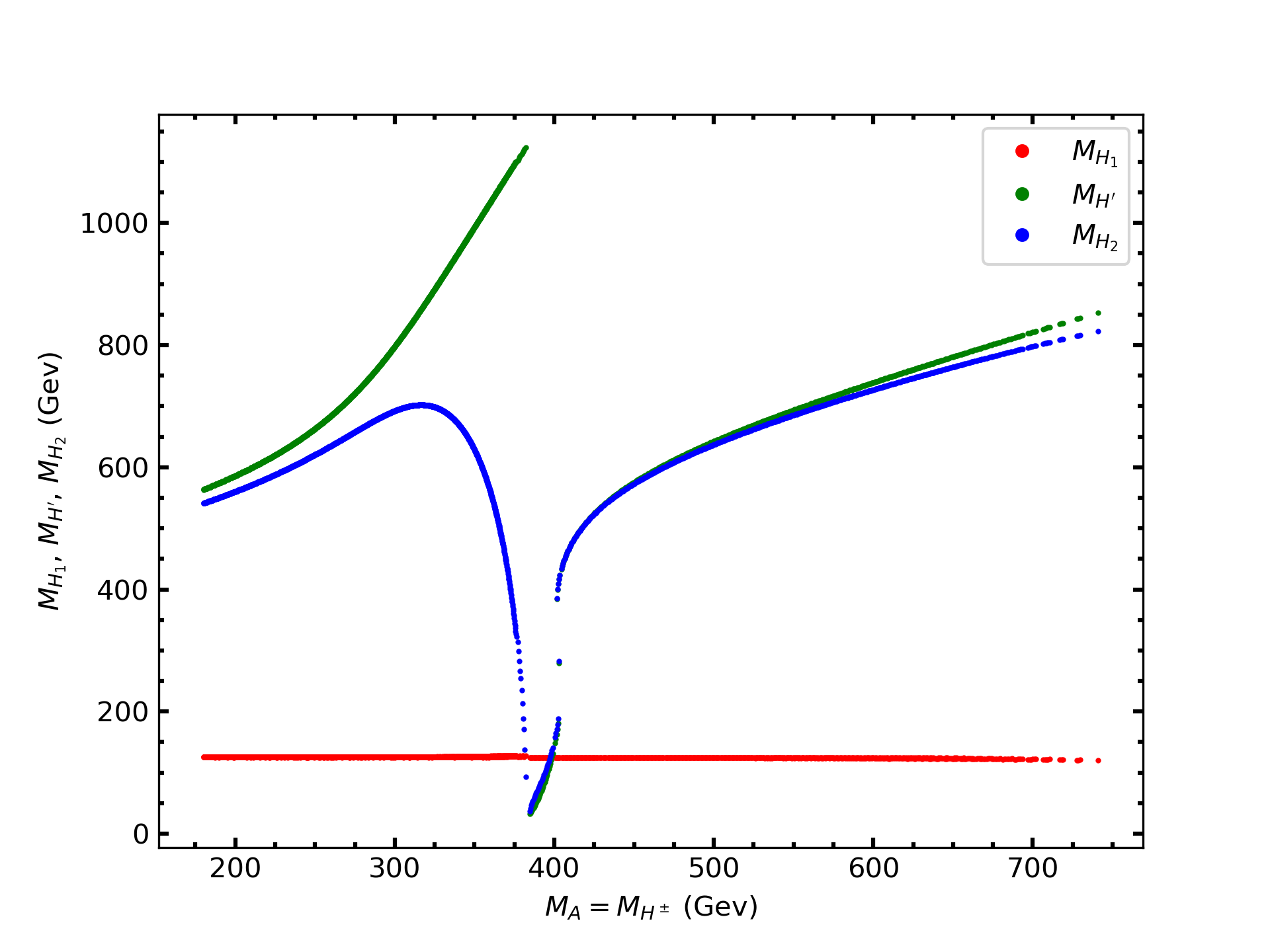}
\includegraphics[width=2.65in,height=2.65in]{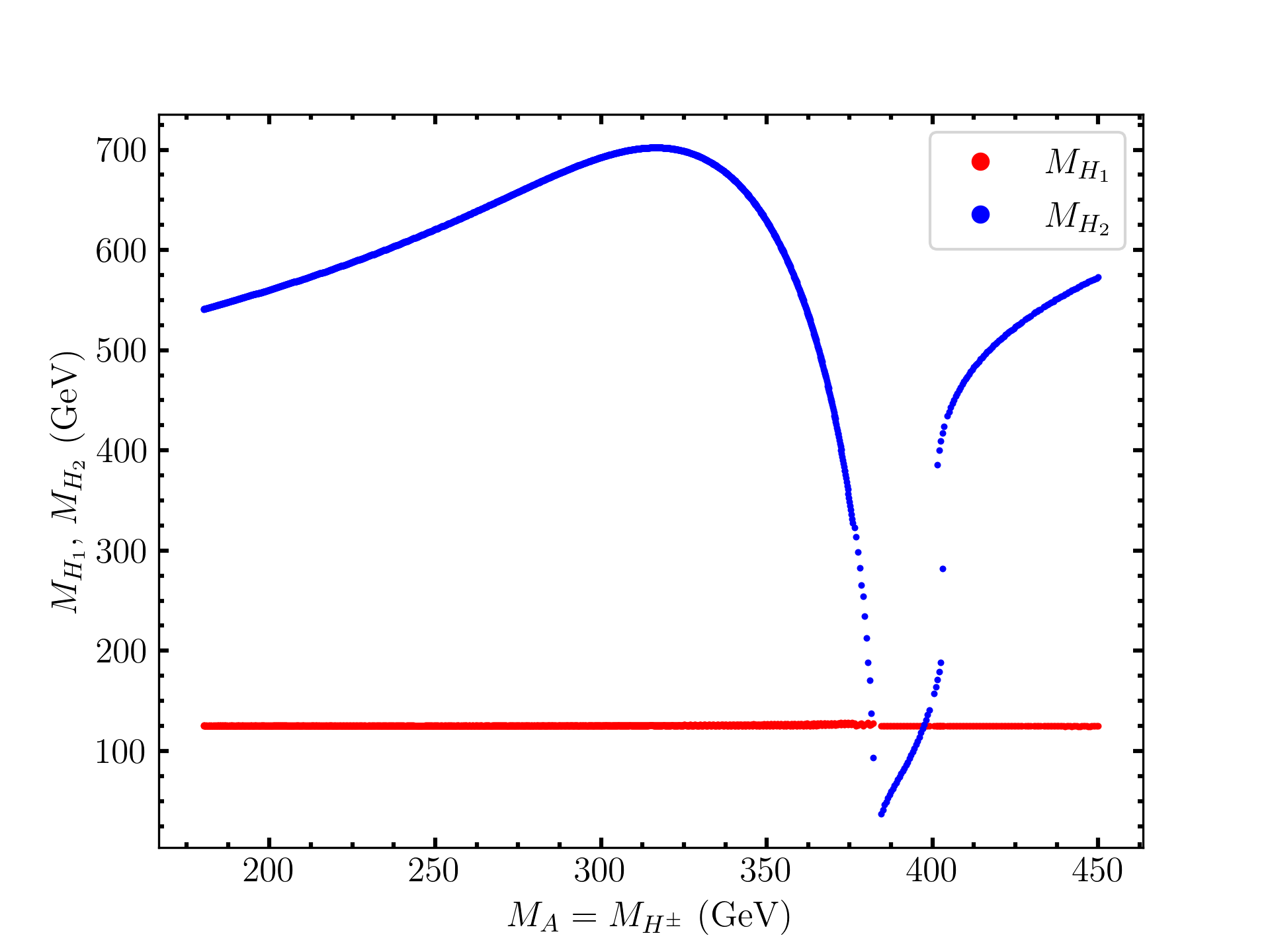}
\caption{Left: The two-loop $C\!P$-even Higgs mass $M_{H_2}$ with $M_{H_1}$
  fixed near $125\,\gev$ as described in the text. The input $M_{H'}$ is
  shown for comparison with $M_{H_2}$. The $B1$--$B2$ transition between the
  two branches of $M_{H_2}$ vs.~$M_A$ occurs at
  $M_A^*({\rm pert.}) \cong 380\,\gev$. Right: A close-up of the $B1$--$B2$
  transition region}
  \label{fig:MH2}
 \end{center}
\end{figure}

Figure~\ref{fig:MH2} shows $M_{H'}$ and the $\CO(\kappa^2)$ $C\!P$-even
eigenmasses $M_{H_1,H_2}$ from $M_A = 180\,\gev$ to $750\,\gev$ in branches
$B1$ and $B2$. While $M_{H'}$ and $M_{H_2}$ start together near $550\,\gev$,
$M_{H'}$ grows to above $1\,\tev$ on branch $B1$, $M_{H_2}$ starts at
$M_{H'} \cong 550\,\gev$ and grows to near $700\,\gev$ at
$M_A \cong 325\,\gev$. It then drops precipitously, falling below $M_{H_1}$
to near zero at $M_A^*({\rm pert.)}$. At that point, the $B2$ branches of
$M_{H'}$ and $M_{H_2}$ emerge and grow rapidly {\em together} from well below
$M_{H_1}$ to about $500\,\gev$ and, then, linearly with and approximately
equal to $M_A$ up to about $750\,\gev$. There is no evidence for a Higgs-like
boson below $100\,\gev$.\footnote{For a more optimistic view, see
  Ref.~\cite{Heinemeyer:2021msz} and references therein.} Also suspicious is
the long linear growth with $M_A$ of $M_{H'}$ and $M_{H_2}$ in $B2$. For
these reasons, we regard branch $B2$ as unphysical.

The behavior of $M_{H_2}$ in branch $B1$ is similar to its one-loop
approximation $\sqrt{(\CM^2_{H'H'})_1} \cong M^2_{H'}$ in
Fig.~\ref{fig:V1masses}. That behavior was caused by Eq.~(\ref{eq:MHsq}) for
$M^2_H$ and its consequence, the sum rule~(\ref{eq:MHsum}) for the BSM Higgs
bosons' masses. That sum rule forced $M_{H'}$ {\em and} the $\CO(V_1)$
eigenmass $M_{H_2}$ to be large when $M_{A,H^\pm}$ were near the experimental
lower bound of $180\,\gev$ for $M_{H^\pm}$ and, then, to plunge to zero when
$(M^4_A + 2M^4_{H^\pm})^{-1/4} \to 540\,\gev$.

A similar thing is happening here: setting $M_{H_1} \cong 125\,\gev$ is a
strong constraint on the BSM Higgs masses, although its mechanics are less
obvious. First, $(\CM^2_{HH})_2$ in Eq.~(\ref{eq:HH}) is dominated by its
first and third terms, the one-loop Higgs mass,
$M^2_H = \sum_n \alpha_n M^4_n/8\pi^2v^2$, and
$(\partial V_2/\partial H^2)_{\langle\,\rangle}$. The condition $M^2_H > 0$
requires
$M^4_{H'} + M^4_A + 2 M^4_{H^\pm} > 12 M^4_t \simeq 10^{10}\,\gev^4\,\,(\gg
6M^4_W + 3M^4_Z)$. This favors large BSM masses, and as $M_{A,H^\pm}$
increase, so does $M_{H'}$ which is being forced to be large by the
algorithm's conditions~(5b,c). Thus, the $M_{H_1}$ constraint requires
$(\partial V_2/\partial H^2)_{\langle\,\rangle} < 0$ and increasing in
magnitude.

The other feature of Fig.~\ref{fig:MH2} in common with the one-loop masses in
Fig.~\ref{fig:V1masses} is $M_{H_2}$ falling from its maximum value to near
zero at the $B1$--$B2$ transition at $M_A^*({\rm pert.})$. The dominant terms
at large BSM masses in Eq.~(\ref{eq:HpHp}) are $M^2_{H'} > 0$ and the last
two, $(-6\cot 2\beta/v) (\partial V_2/\partial H')_{\langle\,\rangle} < 0$
and $(\partial^2 V_2/\partial H^{\prime\,2})_{\langle\,\rangle} > 0$. All
three terms are large and there is a tug-of-war between the latter two which
the negative term wins, driving $M_{H_2} \cong \sqrt{(\CM^2_{H'H'})_2}$ below
$125\,\gev$ --- a level {\em crossing} between the two $C\!P$-even
eigenvalues. This is the same as its behavior in
Figs.~\ref{fig:V1masses}. Recall that, in that figure and this one, the
diagonalization of $\CM^2_{0^+}$ was carried out strictly to $\CO(V_1)$ and
$\CO(V_2)$, respectively, by omitting or truncating the off-diagonal
$\CM^2_{HH'}$ term.\footnote{In Sec.~III, the same dive of $M_{H_2}$ occurs
  at the $B1$--$B2$ transition, but a level {\em repulsion}, not a level
  crossing, occurs there because the full $\CM^2_{HH'}$ is included in
  diagonalizing $\CM^2_{0^+}$ --- as was done in Fig.~\ref{fig:V1angle}.}


The one and two-loop $H$--$H'$ mixing angles are negative and nearly equal to
each other, $\delta_1 \cong \delta_2 \cong -0.001$ for
$180\,\gev < M_A \simle 350\,\gev$. Above $350\,\gev$ $\delta_2$ decreases
rapidly to $-0.028$ at $M_A \cong M_A* = 380\,\gev$ because the denominator
$(\CM^2_{H'H'} - \CM^2_{HH})_2 \cong M^2_{H_2} - M^2_{H_1}$ of
$\tan 2\delta_2$ is becoming smaller as the $B1$--$B2$ transition is
approached; see Fig.~\ref{fig:MH2}.

The experimental consequences of the perturbative method are presented in
Sec.~IV.
 
\section*{III. The GW-2HDM model at two-loops: the amoeba method}

In this section we use a different method to calculate the eigenmasses and
eigenvectors of the $C\!P$-even scalars $H$ and $H'$. The results of this
calculation and the one in Sec.~II are similar. The reasons for this and for
their differences will be explained. In this method, the potential
$V_{\rm eff} = V_0 + V_1 +V_2$ for the GW-2HDM is a function of:
$\tan\beta = v_2/v_1$, the ratio of the VEVs of the $C\!P$-even components of
the complex Higgs doublets $\Phi_2$ and $\Phi_1$; the BSM Higgs masses $M_A$,
$M_{H^\pm}$ and $M_{H'}$;\footnote{Recall that the tree-approximation
  extremal conditions remain in force which, with $\tan\beta$, reduce the
  number of independent quartic couplings to three, namely, $\lambda_5$,
  $\lambda_{45}$ and $\lambda_{345}$. We also remind the reader that an upper
  limit on $\tan\beta$ in this model is a lower limit on $\tan\beta$ in the
  usual 2HDM's with natural flavor conservation~\cite{Branco:2011iw}.}  the
classical fields $H_c$ and $H'_c$ corresponding to the VEVs of the
aligned-basis fields $H$ and $H'$; and the renormalization scale $\LGW$.

We fix $\tan\beta = 0.50$, the experimental upper limit from the searches by
CMS~\cite{Khachatryan:2015qxa} and ATLAS~\cite{Aaboud:2018cwk} for
$gg \to t \bar b H^- \to t \bar t b \bar b$ for
$180\,\gev < M_{H^\pm} \simle 500\,\gev$. We also adopt the precision
electroweak constraint $M_A = M_{H^\pm}$~\cite{Battye:2011jj,
  Pilaftsis:2011ed,Lee:2012jn}, and we assume that
$M_{H',A,H^\pm} \simle \CO(1\,\tev)$, a conservative upper limit suggested by
the analyses of Secs.~I and~II.

 The program {\em Amoeba}~\cite{PresTeukVettFlan92} is used to minimize
 $V_{\rm eff}$ with respect to $H_c$ and $H'_c$ subject to the constraint,
 \be\label{eq:vconstraint}
 H^2_c + H^{\prime\,2}_c = v^2 = (246.2\,\gev)^2,
 \ee
 and with respect to $\LGW$.\footnote{Because of the
   constraint~(\ref{eq:vconstraint}, this minimization involves two
   independent parameters, as in the perturbative method.} This procedure is
 carried out for BSM masses below $1\,\tev$.  Its outputs are the
 renormalization scale $\LGW$, the VEV shift $H'_c$ (with the corresponding
 shift in $H_c$ dictated by Eq.~(\ref{eq:vconstraint})), and the eigenvalues
 $M_{H_1,H_2}$ and eigenvectors $H_1,H_2$ of the $C\!P$-even mass
 matrix. Minimization of $V_{\rm eff}$ requires that $\CM^2_{0^+}$ is a
 positive-definite matrix. This is not yet enough to realistically fix
 $M_{H_1,H_2}$, $H_1,H_2$, and $\LGW$. That happens when we require that one
 of the $C\!P$-even eigenmasses is $M_H = 125\,\gev$. We refer to this
 procedure as the ``amoeba method''.

\begin{figure}[ht!]
   \begin{center}
     \includegraphics[width=2.65in,height=2.70in]{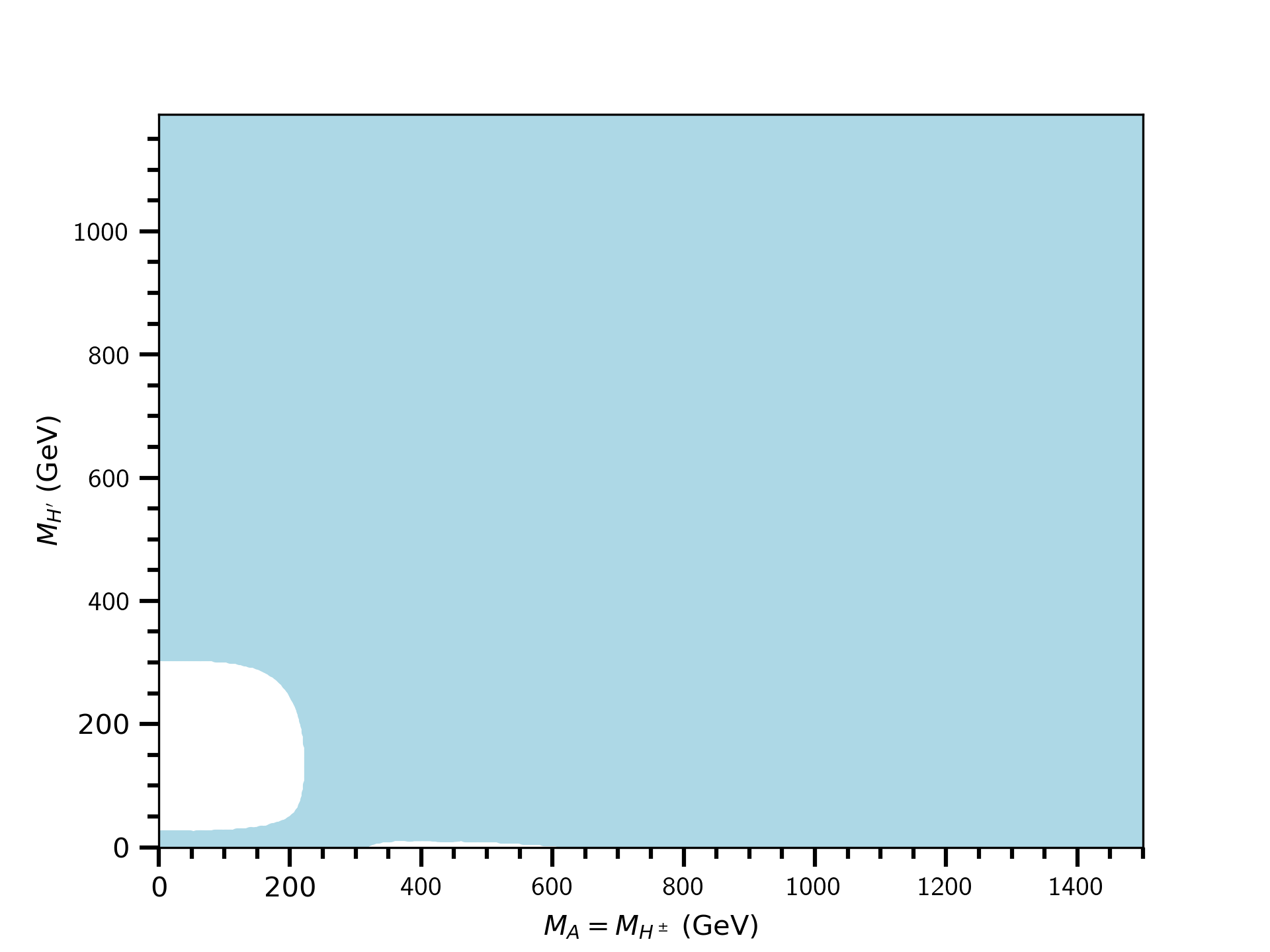}
     \includegraphics[width=2.65in,height=2.70in]{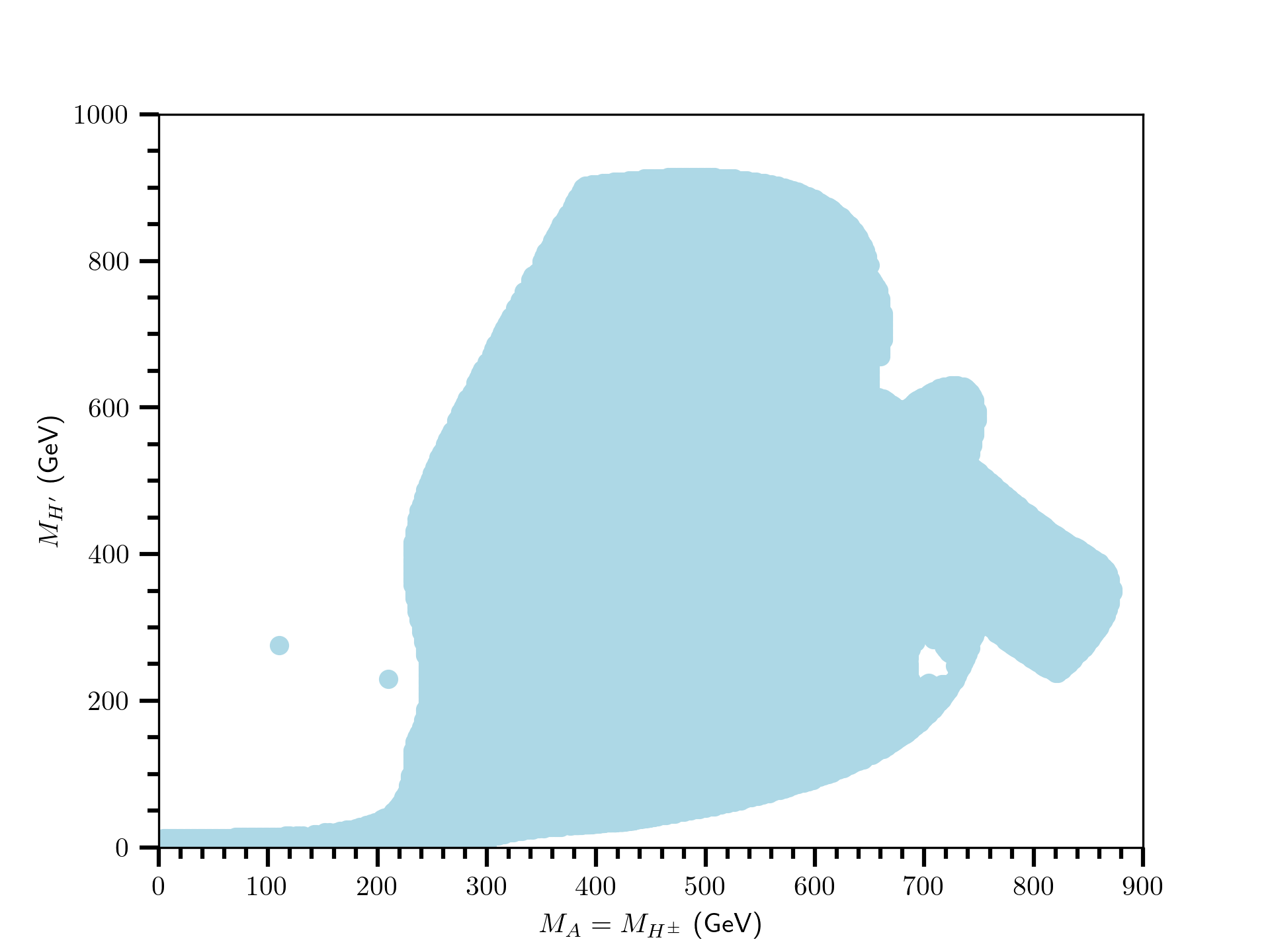}
     \caption{Left: The region (shown in blue) of
       $0 < M_A = M_H^\pm < 1450\,\gev$ and $0 < M_{H'}< 1200\,\gev$ for
       which the one-loop effective potential $V_0 + V_1$ has a minimum as
       described in the text. Right: The Mr.~Magoo region (in blue) for which
       the two-loop potential $V_0 + V_1 + V_2$ has a minimum for the same
       ranges of $M_A = M_{H^\pm}$ and $M_{H'}$. The Higgs mass $M_{H_1}$ has
       not been fixed at $125\,\gev$ in these plots.}
       \label{fig:Vmaps}
   \end{center}
 \end{figure}

 The regions of stability of the one- and two-loop effective potentials for
 BSM Higgs masses below 1~TeV are shown in Figs.~\ref{fig:Vmaps}. Except for
 the small top-quark terms in $V_1$, the one-loop potential is a function of
 $H_c$ and $H'_c$ only through $H^2_c + H^{\prime\,2}_c = v^2$ and, so, it is
 nearly independent of them. This accounts for its large region of stability
 below $1\,\tev$. The small hole near the origin of this plot occurs because
 Eq.~(\ref{eq:MHsq}) cannot be satisfied for $M^2_H > 0$ for that region of
 BSM masses. The cubic and quartic couplings that enter $V_2$ constrain the
 region of stability of the full two-loop potential to
 $300\,\gev \simle M_A = M_{H^\pm} \simle 900\,\gev$ and
 $25\,\gev \simle M_{H'} \simle 900\,\gev$. The mass scale of these ranges is
 set by $v = 246\,\gev$, of course. From now on, we require that {\em one of
   the $C\!P$-even eigenmasses $M_{H_1,H_2} = M_H = 125\,\gev$.} We will see
 that only the case $M_{H_1} = 125\,\gev$ is allowed experimentally.

 Most notably, the amoeba method differs from the perturbative one in that
 $\CM^2_{0^+}$ is required to be positive-definite. Its determinant therefore
 contains terms of order three and four loops ($\CO(\kappa^3)$ and
 $\CO(\kappa^4)$). Furthermore, its eigenvectors and eigenmasses also contain
 terms of $\CO(\kappa^3)$ and $\CO(\kappa^4)$:
\bea
\label{eq:H1H2evecs_Amoeba}
H_1 &=& H\cos\delta - H'\sin\delta,\nn\\
H_2 &=& H\sin\delta + H'\cos\delta,\\
\label{eq:H1H2evals_Amoeba}
M^2_{H_1} &=&
\CM^2_{H_c H_c}\cos^2\delta + \CM^2_{H'_c H'_c}\sin^2\delta -
2\CM^2_{H_c H'_c}\sin\delta \cos\delta,\nn\\
M^2_{H_2} &=&
\CM^2_{H_c H_c}\sin^2\delta + \CM^2_{H'_c H'_c}\cos^2\delta +
2\CM^2_{H_c H'_c}\sin\delta \cos\delta,
\eea
where, now, the $H$--$H'$ mixing angle~$\delta$ is obtained from the ratio
of derivatives with respect to $H_c$ and $H'_c$ of the full two-loop
$V_{\rm eff}$:
\be\label{eq:tandelta_Amoeba}
\tan 2\delta = \frac{2\CM^2_{H_c H'_c}}{\CM^2_{H'_c H'_c} - \CM^2_{H_c H_c}}.
\ee
Since $V_{\rm eff}$ depends on the ``tree-level'' BSM Higgs masses,
$M_A = M_{H^\pm}$ and $M_{H'}$ in Eq.~(\ref{eq:mevec}), this procedure also
determines the allowed ranges of those masses.

\begin{figure}[ht!]
   \begin{center}
     \includegraphics[width=2.65in,height=2.70in]{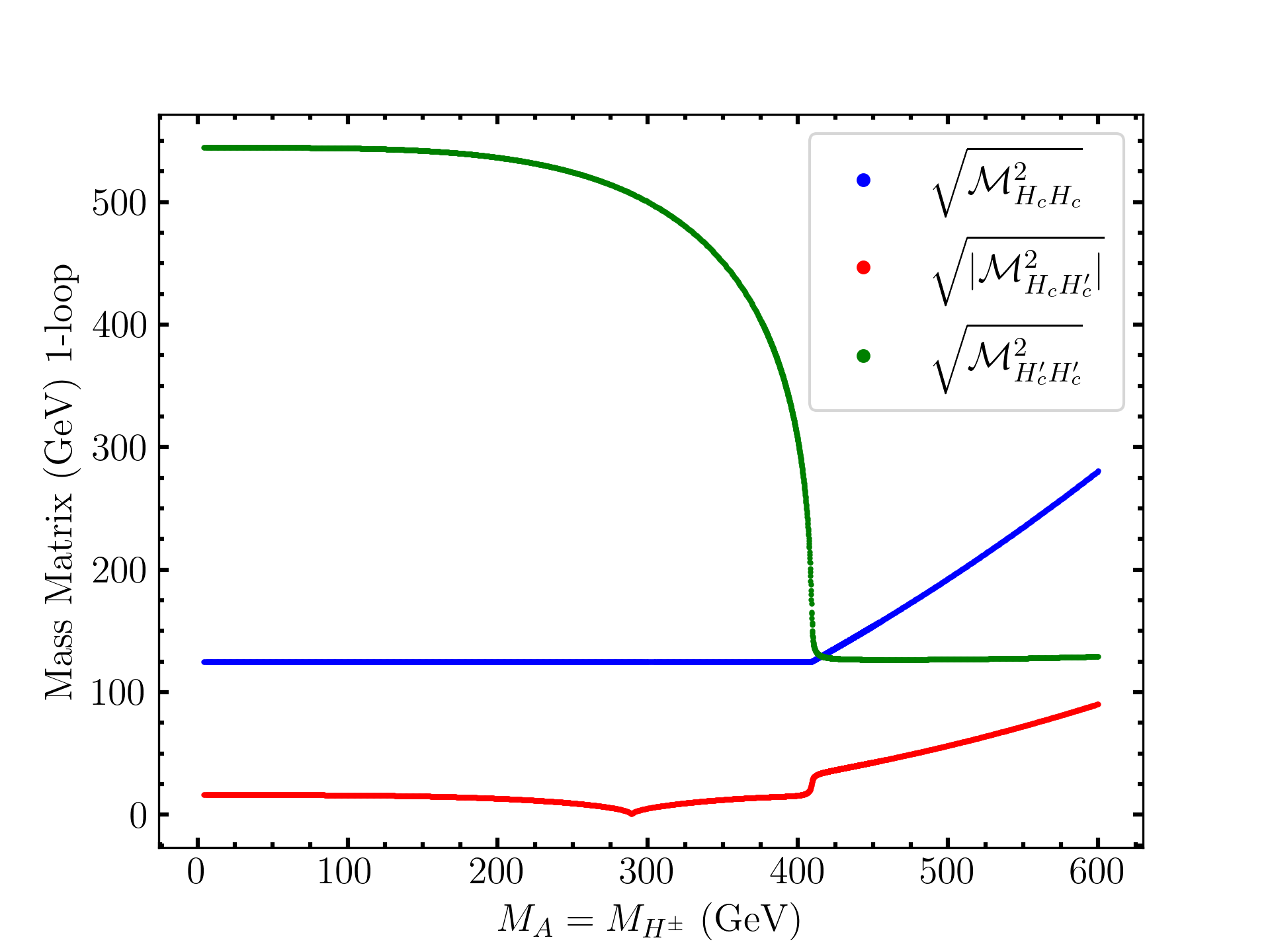}
     \includegraphics[width=2.65in,height=2.70in]{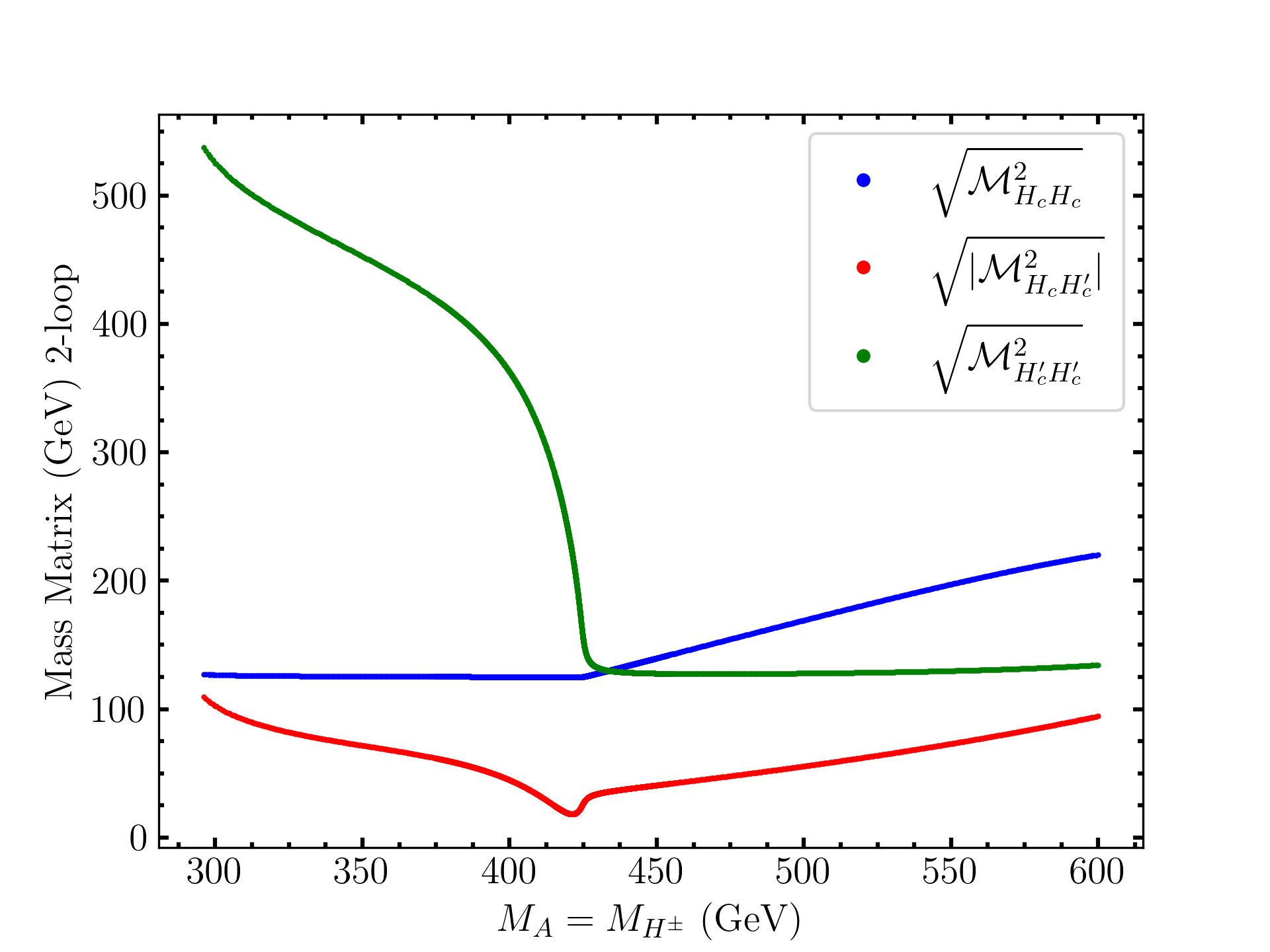}
     \caption{The square roots of the elements of $\CM^2_{0^+}$ in the
       amoeba method for the one-loop effective potential (left) and
       the two-loop effective potential (right). The lighter $C\!P$-even
       eigenvalue is required to be $125\,\gev$ here.}
     \label{fig:M_elts}
   \end{center}
 \end{figure}

\begin{figure}[ht!]
   \begin{center}
     \includegraphics[width=2.65in,height=2.70in]{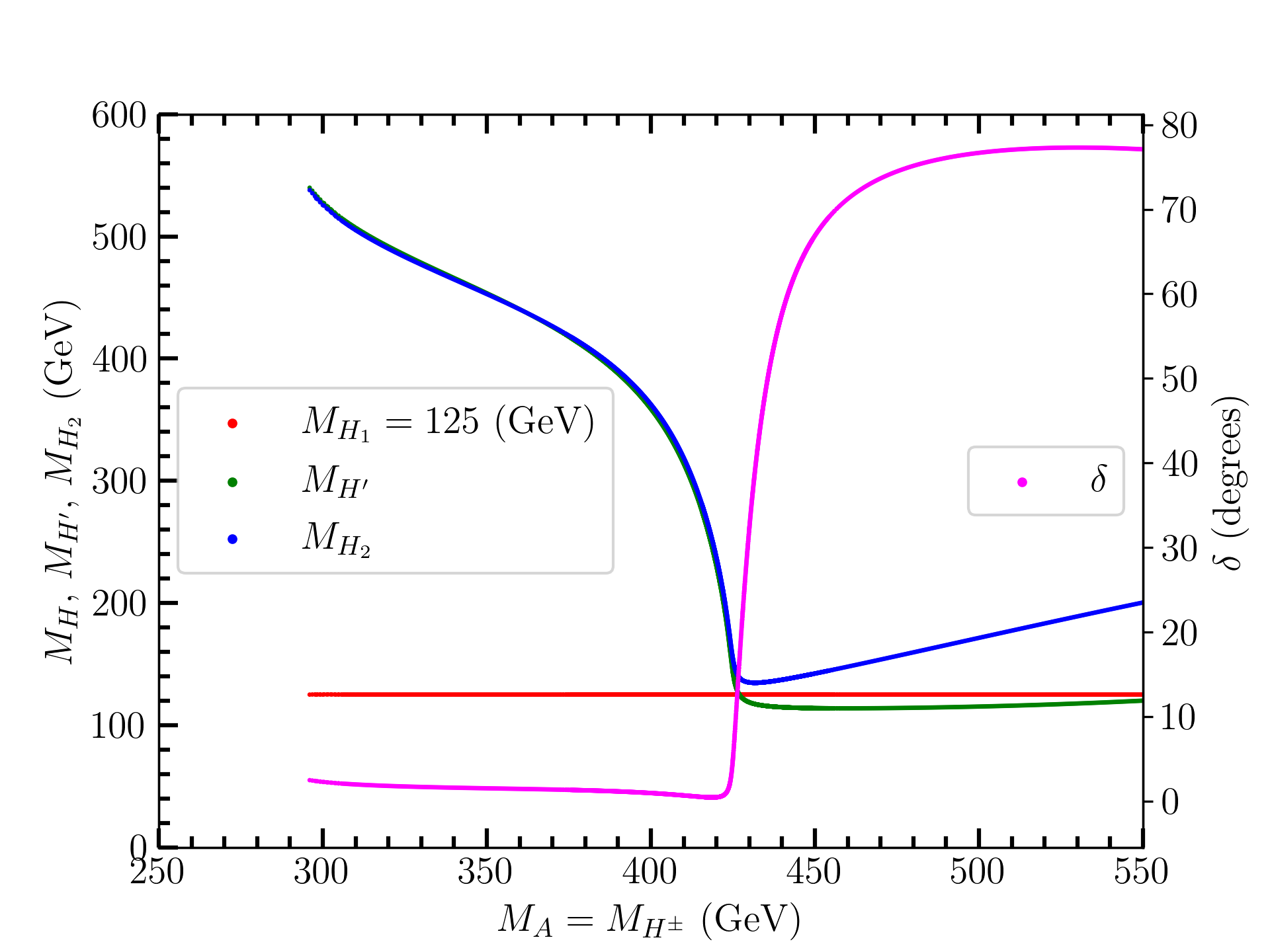}
     \includegraphics[width=2.65in,height=2.70in]{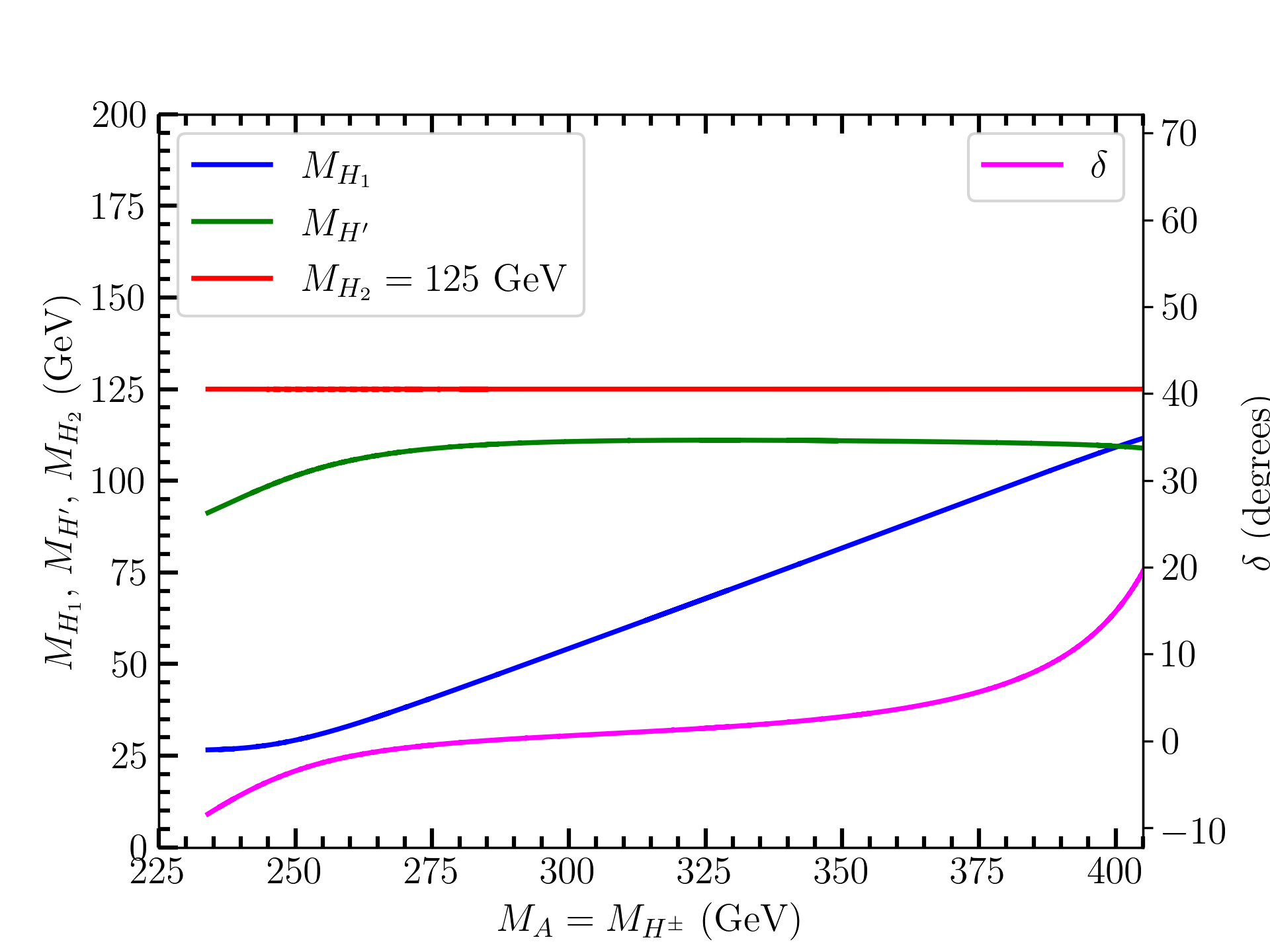}
     \caption{Left: The BSM Higgs masses for the case that the smaller
       $C\!P$-even eigenmass is $M_{H_1} = 125\,\gev$. Also shown (in
       magenta) is the {\em full two-loop} $H$--$H'$ mixing angle (in
       degrees) obtained from Eq.~(\ref{eq:tandelta}). Right: The BSM Higgs
       masses and two-loop $H$--$H'$ mixing angle when the larger $C\!P$-even
       eigenmass is $M_{H_2} = 125\,\gev$.}
     \label{fig:Hlow_high}
   \end{center}
 \end{figure}

As an application of the amoeba method, one that highlights its difference
from the perturbative method of Secs.~I and~II, we apply it to the one-loop
potential $V_0 + V_1$, requiring that the lighter eigenvalue of $\CM^2_{0^+}$
equals $125\,\gev$. The square roots of the elements of the one-loop
$\CM^2_{0^+}$ are shown in the left panel of Fig.~\ref{fig:M_elts}. For
$M_A = M_{H^\pm} \simle 410\,\gev$, note how small the off-diagonal
$\sqrt{|\CM^2_{H_c H'_c}|}$ is compared to the diagonal elements. This is the
hallmark of Higgs alignment in this approximation of the GW-2HDM, in
particular, that $\sqrt{\CM^2_{H_c H_c}} \cong M_{H_1} = 125\,\gev$ and
$\sqrt{\CM^2_{H'_c H'_c}} \cong M_{H_2} > 125\,\gev$. Also, the BSM masses
satisfy the sum rule Eq.~(\ref{eq:MHsum}), which is built into
$V_1$.\footnote{This calculation covered $M_{A,H^\pm} = 0$ to $600\,\gev$,
  endpoints outside the range of the experimental lower bound on $M_{H^\pm}$
  and the one-loop sum rule, but for which $V_{\rm eff}$ has a stable
  minimum.} Here, the important difference with the perturbative method is
the appearance of the small contribution
$\pm \,2 \CM^2_{H_c H'_c}\sin\delta \cos\delta$ in
Eqs.~(\ref{eq:H1H2evals_Amoeba}). This term was excluded in Sec.~I because it
is $\CO(\kappa^2)$. It effects the eigenvalues of $\CM^2_{0^+}$ only very
near $M_A = 410\,\gev$ --- where the denominator of $\tan 2\delta$ is
vanishing. In Fig.~\ref{fig:V1masses}, the region $M_A > 410.5\,\gev$ is
unphysical because it violates the sum rule, and the curves end there. In the
amoeba method, the $\sqrt{\CM^2_{H_c H_c}}$ and $\sqrt{\CM^2_{H'_c H'_c}}$
curves cross at $M_A = 410.5\,\gev$ and $\sqrt{\CM^2_{H_c H_c}}$ rises
approximately linearly while
$\sqrt{\CM^2_{H'_c H'_c}} = 125\,\gev$.\footnote{This is {\em not} a level
  crossing. In fact, the {\em eigenvalues} repel each other there as can be
  seen in Fig.~\ref{fig:Hlow_high} for the two-loop masses.}  As in the
perturbative method, this region is unphysical; i.e., even in one loop of the
amoeba method there are two branches with the transition at the sum-rule
cutoff of $410.5\,\gev$.

\begin{figure}[ht!]
  \begin{center}
    \includegraphics[width = 5.00in,height=3.00in]{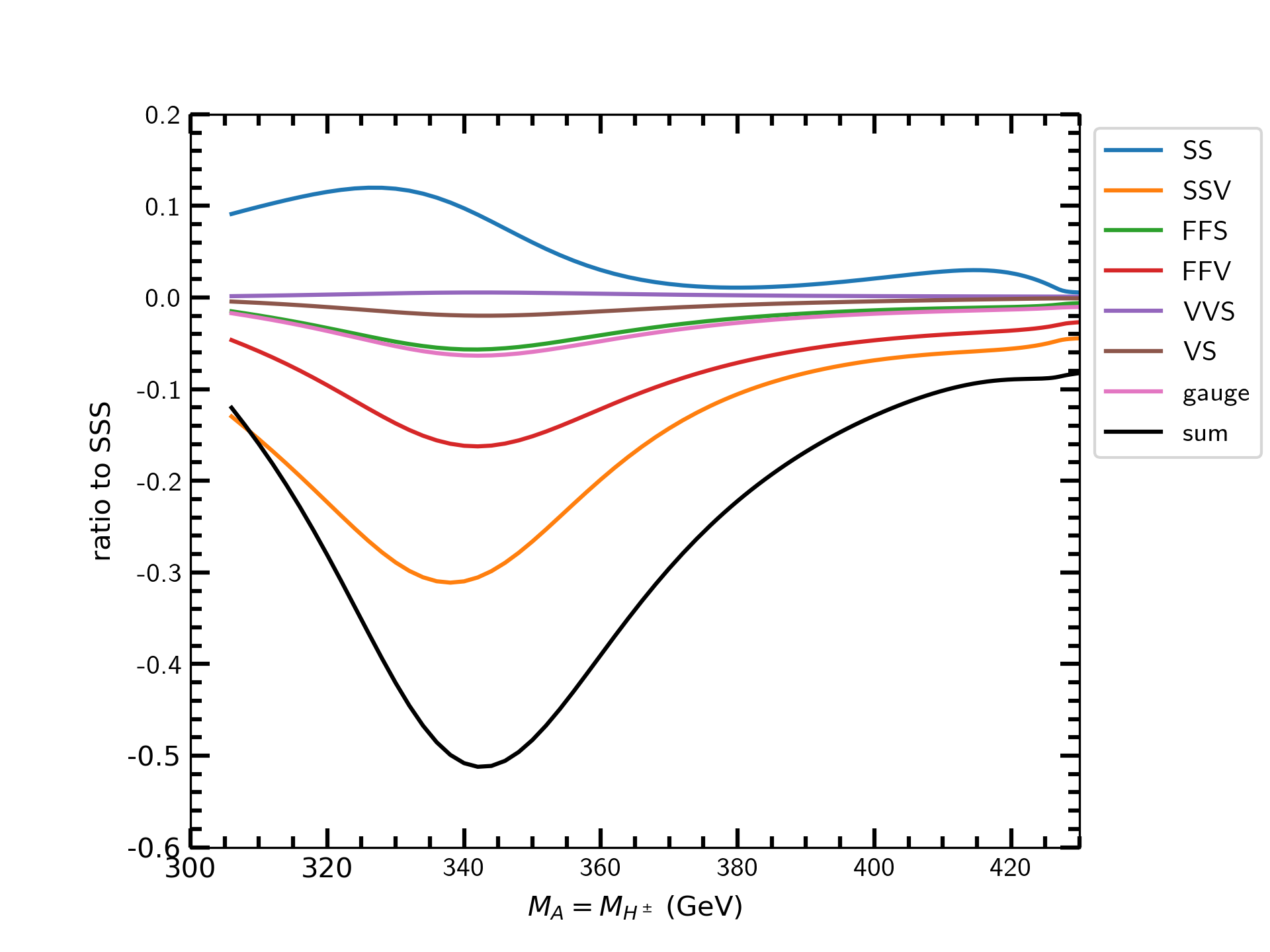}
    \caption{Ratios to the SSS cracked-egg contribution of the SS
      figure-eight, SSV, FFS, FFV, VVS, VS figure-eight, and gauge
      contributions to $V_2$ for
      $290\,\gev \simle M_A = M_{H^\pm} \simle M_A^*({\rm amoe.}) =
      425\,\gev$, the $B1$-branch region of the left panel of
      Fig.~\ref{fig:Hlow_high}. The black curve is the sum of the eight
      ratios. $M_{H_1} = 125\,\gev$ is the lighter $C\!P$-even eigenvalue
      here. In addition to $V_{SSS}$ and $V_{SS}$, these two-loop potentials
      are taken from Martin~\cite{Martin:2001vx}.}
      \label{fig:B1ratios_2}
    \end{center}
  \end{figure}

 Another difference between the two methods is that, in the perturbative one,
 the two-loop effective potential in $B1$ is very well-approximated by its
 all-scalar terms with the cracked-egg (SSS) contribution alone accounting
 for $98\%$ of the total; see Fig.~\ref{fig:B1ratios}. That simplification
 does not occur in the amoeba method. It appears to be due to the different
 regimes of BSM Higgs masses $M_A$ and $M_{H'}$ that give acceptable
 solutions for $M_{H_1,H_2}$ in the two methods. In the perturbative method,
 $M_{H'}$ increases from $550\,\gev$ to $700\,\gev$ and then falls to below
 $125\,\gev$ for $180\,\gev < M_A < M_A^*({\rm pert.}) = 380\,\gev$. In the
 amoeba method, the ranges are $550\,\gev > M_{H'} > 125\,\gev$ for
 $290\,\gev < M_A < M_A^*({\rm amoe.}) = 425\,\gev$. We shall see that the
 region of $V_{\rm eff}$-stability in which the {\em lighter} $C\!P$-even
 eigenvalue $M_{H_1} = 125\,\gev$ will divide into two branches, $B1$ and
 $B2$, with physically acceptable results only in $B1$ --- as in the
 perturbative method. In $B1$, the ratios to the cracked-egg all-scalar
 contribution of several other contributions to the two-loop potential are
 not very small. The sum of the ratios of the other contributions to SSS is
 typically 10--50\%, with the largest contributions after SSS being SSV, FFV
 and SS; see Fig.~\ref{fig:B1ratios_2}. Note that none of these next-largest
 contributions have field-dependent couplings; see Ref.~\cite{Martin:2001vx}
 for details of these potentials.

  \begin{figure}[ht!]
    \begin{center}
     \includegraphics[width=5.30in,height=2.70in]{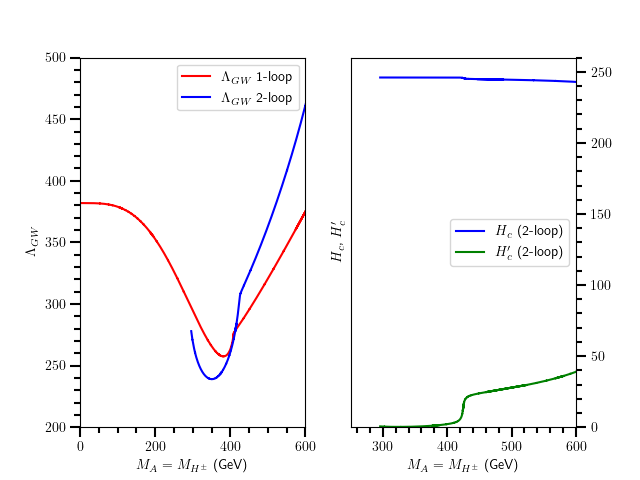}
     \caption{Left:The renormalization scale $\LGW$ in the one- and 2-loop
       approximations of the amoeba method. They are plotted over the ranges
       of the $B1$ and $B2$ branches in the left panel of
       Fig.~\ref{fig:Hlow_high}. Note the slight discontinuity in their
       slopes at the transition between the two branches at
       $425\,\gev$. Right: The classical-field shifts $H_c$ and $H'_c$ at the
       minima of the two-loop $V_{\rm eff}$ as a function of
       $M_A = M_{H^\pm}$. The smallness of $H'_c$, especially in branch $B1$,
       is a consequence of the requirement that $\CM^2_{H_c H'_c}$ is small
       enough that $\det{\CM^2_{0^+}} > 0$.}
     \label{fig:align_ee}
   \end{center}
 \end{figure}

 The left panel of Fig.~\ref{fig:align_ee} shows the one- and two-loop
 renormalization scale $\LGW$ over the ranges of the $\sqrt{\CM^2_{0^+}}$ in
 Fig.~\ref{fig:M_elts}; $M_{H_1} = 125\,\gev$ is the lighter $C\!P$-even
 eigenvalue in both curves. The magnitudes of these renormalization scales
 are comparable to those of the two-loop scales $\Lambda_0$ and $\LGW$ in the
 perturbative method (Fig.~\ref{fig:MHpr_LGW}, right panel), their values
 again being set by $v = 246\,\gev$. But, while the renormalization scales in
 the perturbative method are discontinuous at the $B1$--$B2$ transition (as
 is $M_{H'}$), the transitions in the amoeba method are continuous (again, as
 is $M_{H'}$ in the left panel of Fig.~\ref{fig:Hlow_high}), albeit with
 slight changes of slope.

 The right panel of Fig.~\ref{fig:align_ee} is more interesting:
 $H_c \cong 246\,\gev$ in $B1 = 290$--$425\,\gev$ and decreasing only
 slightly in $B2 = 425$--$600\,\gev$; $H'_c$ is negligibly small in $B1$
 (alignment again), but jumps to $20\,\gev$ at the transition and increases
 with $M_A$ from there up to $M_A = 600\,\gev$. Stable solutions of
 $V_{\rm eff}$ are scarce beyond this upper limit of $B2$.

 Finally, we extract the $C\!P$-even masses and states. The two possibilities
 for which eigenmass is $125\,\gev$ are shown in
 Figs.~\ref{fig:Hlow_high}. The masses $M_{H'}$, $M_{H_1}$, $M_{H_2}$ and the
 complete two-loop $H$--$H'$ mixing angle $\delta$ are plotted
 vs.~$M_A = M_{H^\pm}$ for the regions in which $V_{\rm eff}$ has a stable
 minimum.\footnote{When $M_{H_2} = 125\,\gev$, it is difficult to obtain a
   stable minimum of $V_{\rm eff}$ above $M_A \cong 400\,\gev$.}  Clearly,
 only the case that $M_{H_1} = 125\,\gev$ is consistent with light
 Higgs-boson searches from LEP and LHC. In the left panel, $M_{H_2}$ and
 $M_{H'}$ decrease from $550\,\gev$ to just above and below
 $M_{H_1} = 125\,\gev$, and they are indistinguishable up to
 $M_A \cong M_A^*({\rm amoe.}) = 425\,\gev$. As the right panel of
 Fig.\ref{fig:M_elts} and this figure illustrate, this is due to the
 smallness of
 $\delta \cong \CM^2_{H_c H'_c}/(\CM^2_{H'_c H'_c} - \CM^2_{H_c H_c})$ for
 $M_A < M_A^*({\rm amoe.})$. At $M_A^*({\rm amoe.})$, where $\delta$ passes
 rapidly from near zero to $\pi/4$, there is a level repulsion between
 $M_{H_2}$ and $M_{H_1}$.\footnote{This is the same behavior as the
   one-loop eigenmasses and the $H$--$H'$ mixing angle in the right
   panel of Fig.~\ref{fig:V1angle}.}  Beyond that point, $M_{H_2}$ rises
 linearly with $M_A$ {\em and} $M_{H'}$ crosses $M_{H_1}$ but remains nearly
 equal to it. Furthermore, the large value of $\delta$ violates loop
 perturbation theory. As in the perturbative method for the two-loop
 potential, there are two branches of $M_{H_2}$, the physical one $B1$ below
 $M_A^*({\rm amoe.})$ and the unphysical one $B2$ above it.

 It is interesting to compare the masses in the perturbative method,
 Fig.~\ref{fig:MH2}, with those here in the amoeba method. The behaviors of
 $M_{H'}$ in the two methods are radically different, increasing rapidly in
 both branches with a jump discontinuity at the transition in the first
 method, while decreasing to $M_{H_1} = 125\,\gev$ in the second. On the
 other hand, the behaviors of $M_{H_2}$ in the two methods are strikingly
 similar. In $B1$, it starts near $550\,\gev$, its maximum value in the
 second method, not far below its maximum of $700\,\gev$ in the first
 one. Then, in both methods, it dives to well below or just below
 $M_{H_1} = 125\,\gev$ at the $B1$--$B2$ transition. In $B2$, $M_{H_2}$ grows
 linearly with $M_A = M_{H^\pm}$ in both methods. In the perturbative method
 calculation, $M_A$ runs over the range $180$--$1100\,\gev$, but it is clear
 in Fig.~\ref{fig:MH2} that the criteria for generating $M_{H_1}$ and
 $M_{H_2}$ are difficult to meet above $M_A = 700\,\gev$. This is similar to
 the upper limit $M_A = 600$--$700\,\gev$ in the region of stability of the
 two-loop potential in Fig.~\ref{fig:Vmaps}.

 \section*{IV. Experimental consequences for the GW-2HDM in
   two loops}

 We have stressed that the only feasible way of testing the GW-2HDM (and
 similar GW models) in the foreseeable future is to discover or exclude its
 BSM Higgs bosons $H_2$, $A$, $H^\pm$.\footnote{In this section, we use $H$
   and $H_1$ interchangeably because, as we saw in the amoeba method, only
   the lightest $C\!P$-even eigenvalue $M_{H_1} = 125\,\gev$ is consistent
   with Higgs boson searches near and below that mass. We do not use $H'$ and
   $H_2$ interchangeably because of their very different dependence on
   $M_{A,H^\pm}$ in the perturbative method.} As in the one-loop analysis,
 the overriding features of these bosons are (1)~their low masses, well below
 1~TeV, and (2)~the high degree of alignment of the 125~GeV Higgs boson $H$
 and the related strong suppression of the BSM bosons' couplings to $W^+W^-$,
 $ZZ$, $W^\pm Z$ and $ZH$, $W^\pm H$. Additional suppression in their
 production rates is due to the appearance in their Yukawa couplings of
 $\tan\beta \simle 0.50$ for $M_{H^\pm} \simle 500\,\gev$. This section
 summarizes the BSM Higgs mass ranges found in our two-loop calculations,
 their couplings to electroweak gauge bosons and to quarks and leptons, and
 the searches we believe are likely to reveal, or exclude, the BSM
 Higgses.\footnote{Earlier discussions of the BSM Higgs searches for masses
   in the one-loop approximation are in Refs.~\cite{Lane:2018ycs,
     Lane:2019dbc, Eichten:2021qbm}.}

 The BSM masses obtained in our two-loop study are qualitatively similar to
 those found using the simple one-loop sum rule in Eq.~(\ref{eq:MHsum}) and,
 as we have discussed, for much the same reason, namely, the constraint on
 these masses from the requirement that $M_H = 125\,\gev$. As explained in
 Sec.~I, we require $M_{H^\pm} \ge 180\,\gev$ and $M_A = M_{H^\pm}$.  Then,
 the physical (branch $B1$) BSM mass ranges are (from
 Figs.~\ref{fig:MH2},\ref{fig:Hlow_high}):
\bea
\label{eq:pert_masses}
&180\,\gev& \simle M_A = M_{H^\pm} \simle 380\,\gev, \nn\\
&700\,\gev& \simge M_{H_2} \simge 125\,\gev \,\,\,\,{\rm( perturbative\,\,
  method)};\\\nn \\
\label{eq:amoeba_masses}
&290\,\gev& \simle M_A = M_{H^\pm} \simle 425\,\gev, \nn\\
&550\,\gev& \simge M_{H_2} \simge 125\,\gev \,\,\,\,{\rm (amoeba\,\,method)}.
\eea
These ranges are correlated, with $M_{H_2}$ decreasing as $M_A = M_{H^\pm}$
increase. As in the one-loop analysis, $M_{H_2}$ decreases to unrealistically
small values as $M_A = M_{H^\pm}$ increase to their maximum allowed ($M^*_A$)
by the method used.

\begin{figure}[t!]
 \begin{center}
\includegraphics[width=2.675in, height=2.675in]{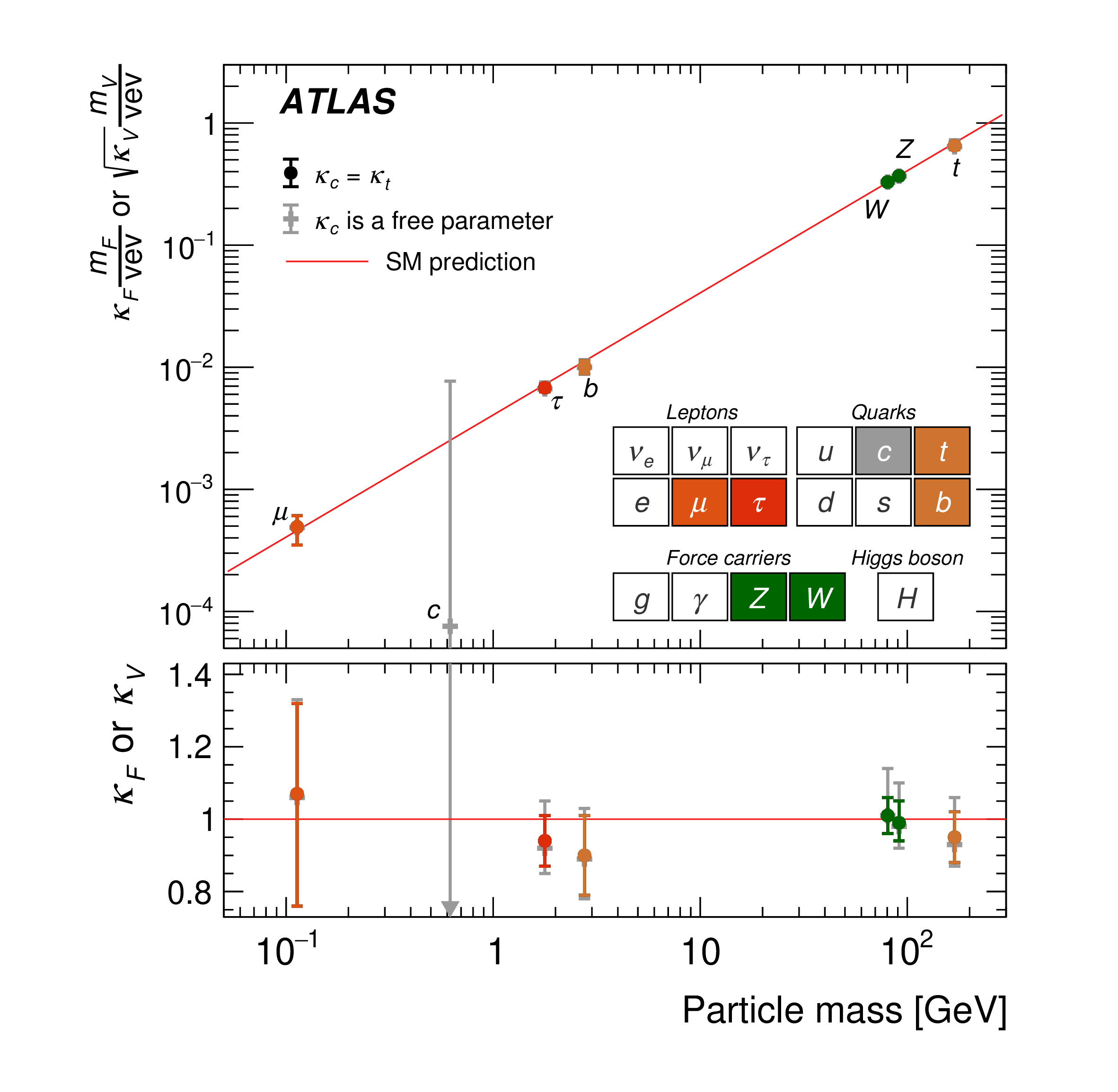}
\includegraphics[width=2.65in, height=2.65in]{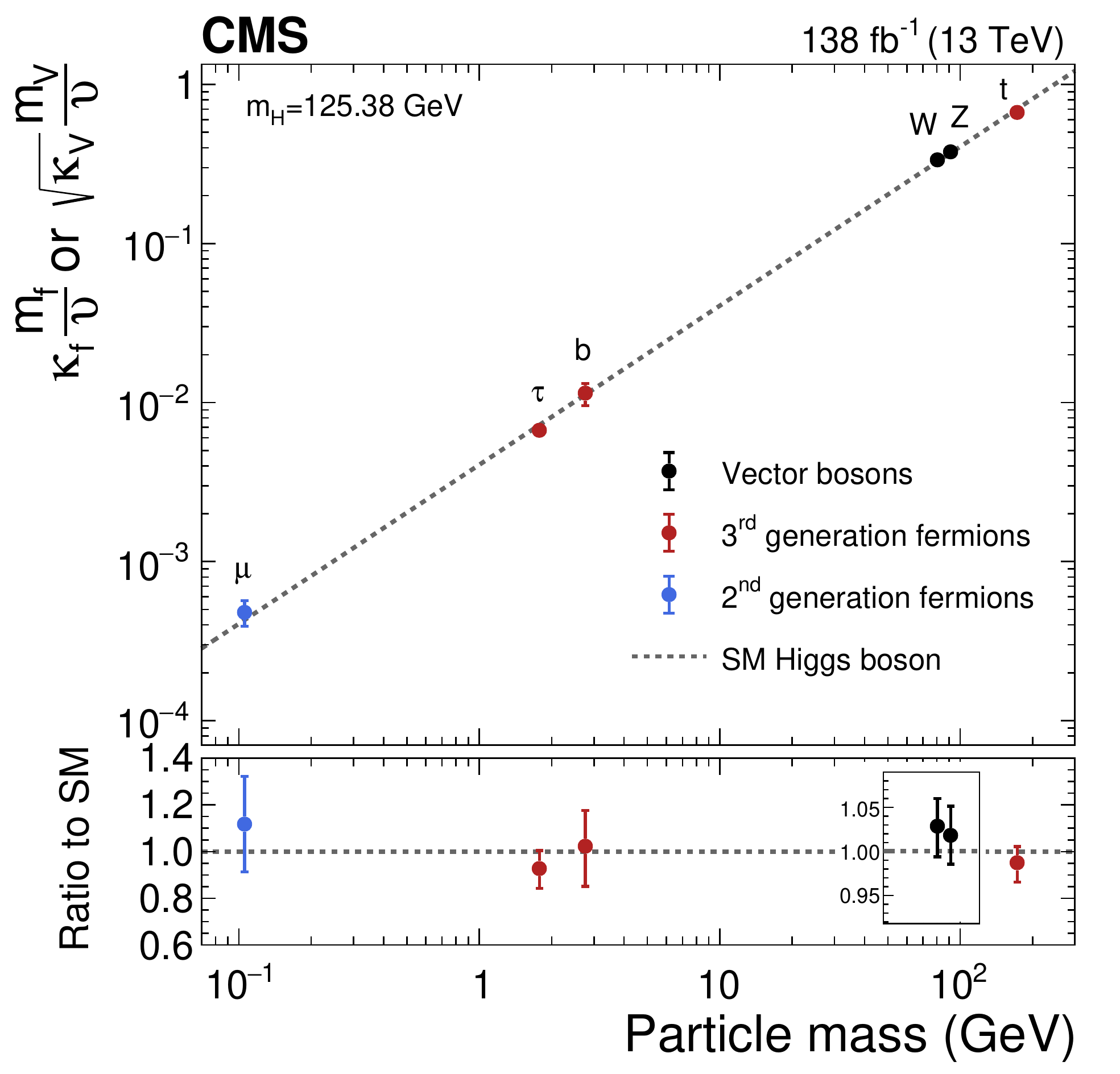}
\caption{The mass-dependent couplings of the 125-GeV Higgs boson $H$ to
  quarks, leptons and the $W$ and $Z$ determined by
  ATLAS~\cite{ATLAS:2022vkf} and CMS~\cite{CMS:2022dwd} from their full LHC
  Run~2 data sets. The lower panel in each figure is the ratio of the
  measured couplings to the Standard Model ones.}
  \label{fig:SMHiggs}
 \end{center}
 \end{figure}

 The degree of Higgs alignment is dramatically illustrated in
 Figs.~\ref{fig:SMHiggs}. These are the full Run~2 determinations of the
 couplings of $H$~to quarks, leptons and weak bosons from
 ATLAS~\cite{ATLAS:2022vkf} and CMS~\cite{CMS:2022dwd}. All the measurements
 are within one standard deviation of the Standard Model prediction. From a
 theoretical point of view, the 125~GeV Higgs boson is either the lone
 ``Higgs'' of the Standard Model~\cite{Glashow:1961tr,Weinberg:1967tq,
   Englert:1964et,Higgs:1964ia,Guralnik:1964eu,Aad:2012tfa,
   Chatrchyan:2012ufa} or Higgs alignment~\cite{Boudjema:2001ii,
   Gunion:2002zf,Carena:2013ooa} is verified experimentally.

 The allowed and strongly suppressed couplings in the GW-2HDM, are in the
 interaction $\CL_{EW}$ of the Higgs bosons with the electroweak gauge
 bosons~\cite{Lane:2018ycs}. Having found that the $H$--$H'$ mixing angle
 $\delta \simle \CO(10^{-2})$ through two-loop order, an excellent
 approximation to $\CL_{EW}$ is obtained by putting $\sin\delta = 0$,
 $H = H_1$ and $H' = H_2$:
\bea\label{eq:PhiEW}
\CL_{EW} &=&  ie H^- \overleftrightarrow{\partial_\mu} H^+
              \left(A^\mu + Z^\mu \cot 2\theta_W \right)
+ \frac{e}{\sin 2\theta_W} \left(H_2\overleftrightarrow{\partial_\mu}A \right)
Z^\mu \nn\\
&+& \frac{ig}{2}\left(H^+
  \overleftrightarrow{\partial_\mu}(H_2+iA) W^{-\,\mu} -
  H^- \overleftrightarrow{\partial_\mu}(H_2 - iA)W^{+\,\mu}\right) \nn\\
&+& H_1 \left(gM_W\, W^{+\,\mu} W^{-}_\mu +
       {\thalf}\sqrt{g^2 + g^{\prime\,2}}M_Z\, Z^\mu Z_\mu\right) + \nn\\
&+& \left(H_1^2+H_2^2 +A^2\right)\left({\tfourth}g^2\, W^{+\,\mu}
W^{-}_\mu +{\teighth}(g^2+g^{\prime\,2})\,Z^\mu Z_\mu\right)\nn\\
&+& H^+H^- \left(e^2(A_\mu + Z_\mu\cot 2\theta_W)^2 +
{\tfourth}g^2 \,W_\mu^+ W^{-\,\mu}\right),
\eea
where $\tan\theta_W = g'/g$ and $e = gg'/\sqrt{g^2 + g^{\prime\,2}}$. The
negative results of LHC searches for the 2HDM Higgs bosons $H_2$, $A$ and
$H^\pm$ are entirely consistent with Eq.~(\ref{eq:PhiEW}); see
{\url{https://twiki.cern.ch/twiki/bin/view/AtlasPublic}} and
{\url{https://cms-results.web.cern.ch/cms-results/public-results/publications}}.

\begin{figure}[t!]
\includegraphics[width=1.0\textwidth]{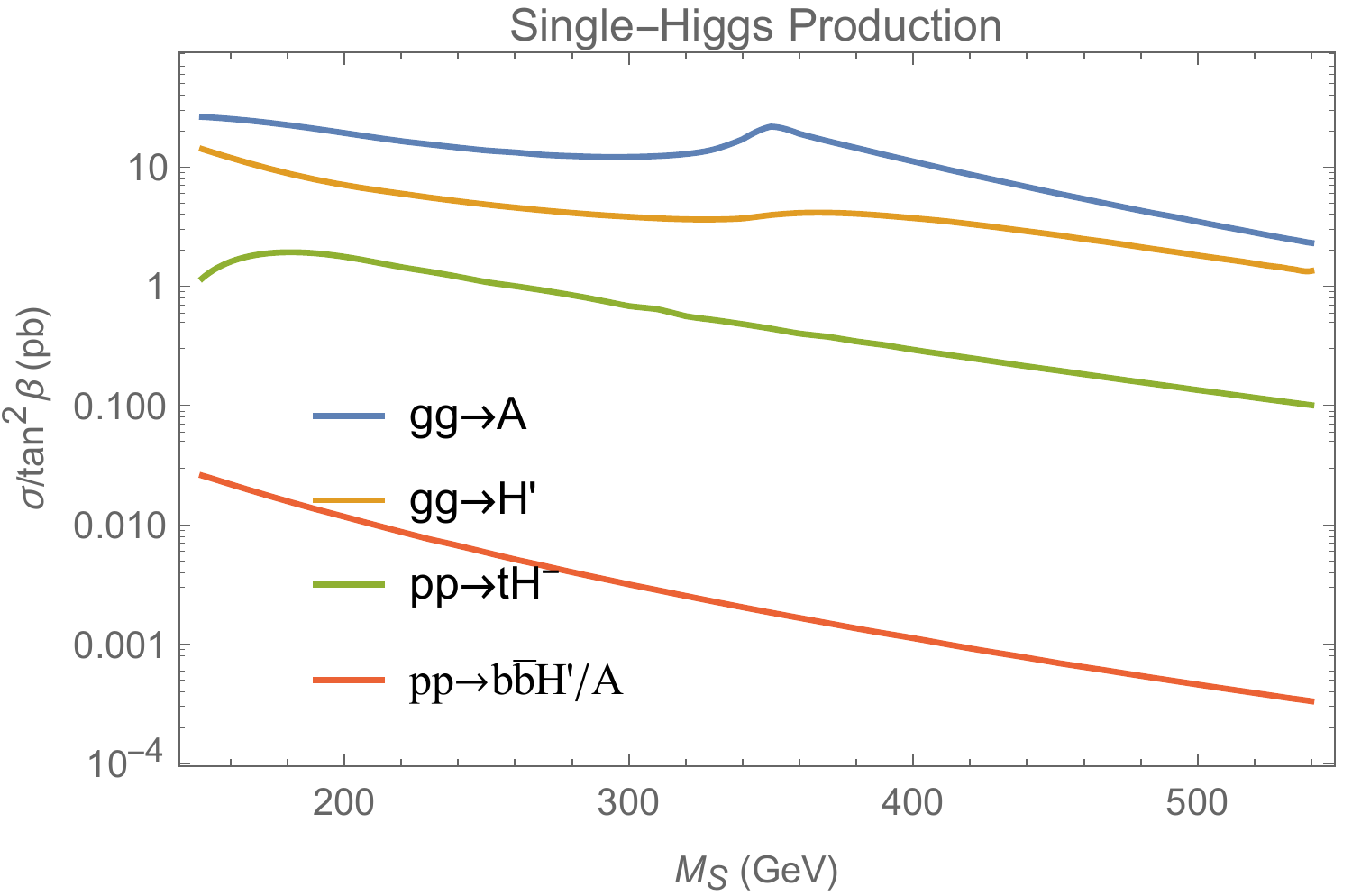}
\caption{The gluon fusion cross sections for $\sqrt{s} = 13\,\tev$ at the LHC
  for single BSM Higgs production in the alignment limit ($\delta \to 0$) of
  the GW-2HDM~\cite{Lane:2018ycs}. The dependence on $\tan\beta$ has been
  scaled out; both charged Higgs states are included in
  $pp \to t\bar b H^-$.}
\label{fig:ggf}
\end{figure}

\begin{figure}[h!]
\includegraphics[width=1.0\textwidth]{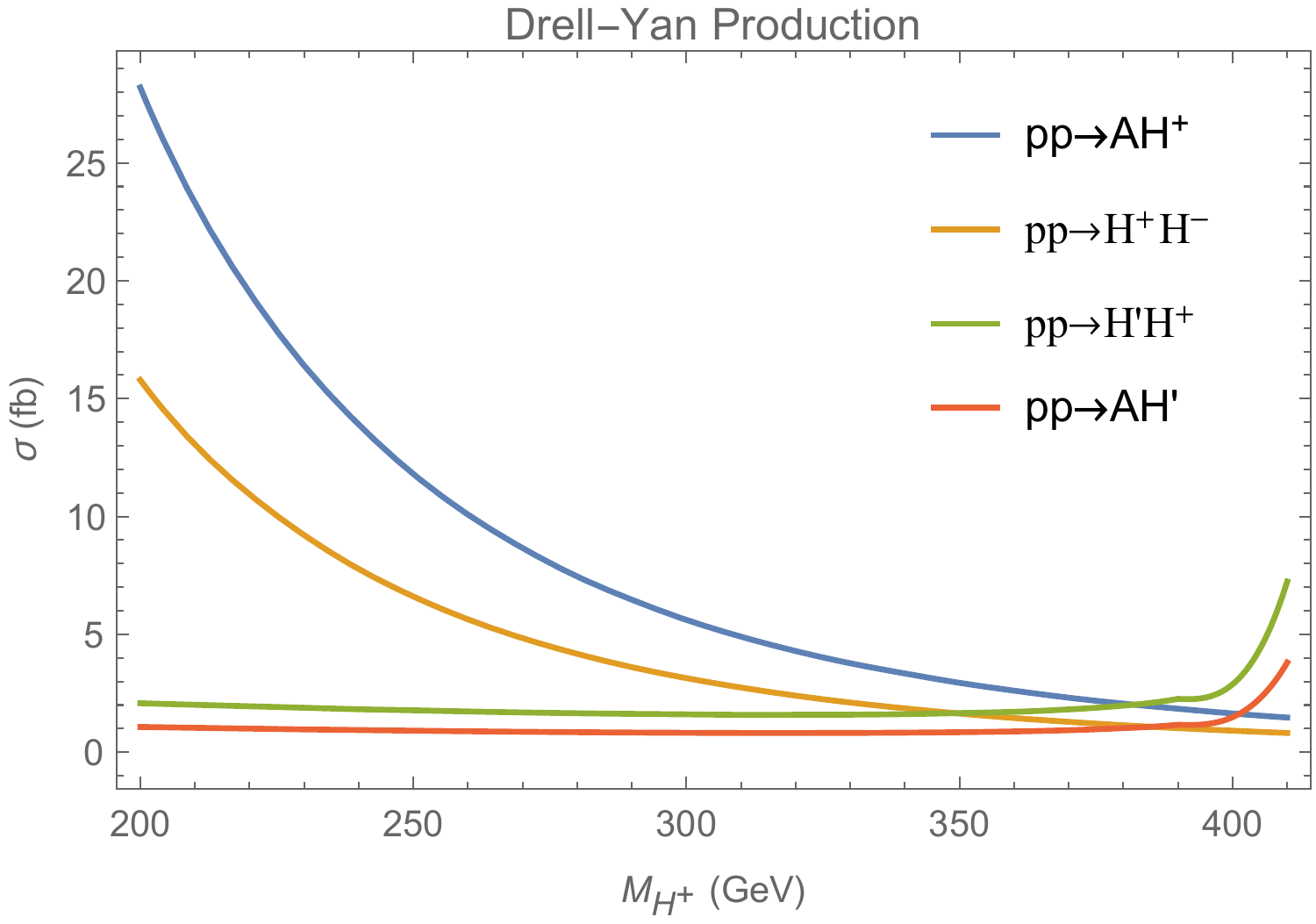}
\caption{The Drell-Yan cross sections for $\sqrt{s} = 13\,\tev$ at the LHC
  for production of Higgs pairs in the alignment limit
  ($\delta \to 0$)~\cite{Lane:2018ycs}. They are independent of
  $\tan\beta$. $M_{H^\pm} = M_A$ is assumed, with $M_{H_2}$ taken from
  Eq.(\ref{eq:MHsum}). The sharp increase at large $M_{H^\pm}$ is due to the
  rapid decrease of $M_{H_2}$ there.}
\label{fig:DY}
\end{figure}

Another, less dramatic but possibly important, suppression due to
$\tan\beta \simle 0.50$ is in the fermions' Yukawa interaction. Because of
alignment and our choice in Eq.~(\ref{eq:Z2}) of the type-I model for the
GW-2HDM, all the BSM Higgs couplings to quarks and leptons are proportional
to $\tan\beta$:\footnote{Hence the upper limit $\tan\beta \simle 0.50$ from
  the LHC searches for $H^\pm$. In the conventional definition of the type-I
  2HDM~\cite{Branco:2011iw}, the analysis in Ref.~\cite{Lane:2018ycs} would
  have found $\cot\beta \simle 0.50$, in significant contradiction, e.g.,
  with the experimental limits~\cite{CMS:2019rvj,ATLAS:2022ohr} discussed
  below.}
\bea\label{eq:yukawa}
\CL_Y &=& \frac{\sqrt{2}\tan\beta}{v}
      \sum_{k,l=1}^3\left[H^+\left(\bar u_{kL} V_{kl}\,m_{d_l}d_{lR}
      -\bar u_{kR}\, m_{u_k} V_{kl}\, d_{lL} +
      m_{\ell_k}\bar\nu_{kL}\ell_{kR}\,\delta_{kl}\right) + {\rm
      h.c.}\right] \nonumber\\
   &-& \left(\frac{v + H_1 - H_2\tan\beta}{v} \right)
       \sum_{k=1}^3 \left(m_{u_k} \bar u_k u_k + m_{d_k} \bar d_k d_k
                         +m_{\ell_k}\bar\ell_k \ell_k\right) \nonumber\\
   &-& \frac{i A\tan\beta}{v} \sum_{k=1}^3 \left(m_{u_k} \bar u_k \gamma_5 u_k
     - m_{d_k} \bar d_k\gamma_5 d_k - m_{\ell_k}\bar\ell_k
     \gamma_5\ell_k\right),
\eea
where $V = U_L^\dagg D_L$ is the CKM matrix.
The cross sections for gluon fusion (with $\tan^2\beta$ scaled out) and
for Drell-Yan production of the BSM bosons at the 13~TeV LHC are shown in
Figs.~\ref{fig:ggf} and~\ref{fig:DY}. Except at low $M_A = M_{H^\pm}$ or
$\tan\beta \simle 0.1$, the gluon fusion rates are typically $\simge 100$
times larger than Drell-Yan ones.

Thus, the most common production processes of the BSM scalars in the GW-2HDM
are:
\bea
\label{eq:ggHpm}
gg &\to& b\bar b \to t\bar b H^- + {\rm c.c.} ,\\
\label{eq:ggHpA}
gg &\to& t \bar t \to H_2,A.
\eea
The process~(\ref{eq:ggHpA}) may go through a top-quark loop or via on-shell
tops with four top quarks in the final state if $M_{H_2}$ or $M_A > 2m_t$;
both possibilities are discussed below. The rate for the common search mode
$gg \to H_2,A \to \gamma\gamma$ via a top loop is suppressed by $\tan^2\beta$
as well as having the usual small $\CO(\alpha^2)$ branching ratio.

For the two-loop mass ranges in Eqs.~(\ref{eq:pert_masses},
\ref{eq:amoeba_masses}) the major BSM decay modes are:\footnote{The
  assumption $M_{H^\pm} = M_A$ precludes $A \to W^\pm H^\mp$.}
\bea
\label{eq:Hpmmodes}
&&H^+ \to t \bar b;\\
\label{eq:Amodes}
&&A \to b \bar b,\,\,\,\tau^+\tau^-,\,\,\,t \bar t;\\
\label{eq:H2modes}
&&H_2 \to b \bar b,\,\,\,\tau^+\tau^-, \,\,t \bar t \,\,{\rm and}\,\,ZA,\,\,W^\pm
H^\mp.
\eea
Since $M_A = M_{H^\pm}$ must be about $100\,\gev$ greater than $M_{H_2}$ to
enable the decays $H^\pm \to W^\pm H_2$ and $A \to ZH_2$, they are forbidden
in the two-loop mass ranges found with the perturbative method. In the amoeba
method, these decays are allowed only for $M_{A,H^\pm} = 410$--$425\,\gev$,
with rates much smaller than $H^+ \to t \bar b$ and
$A \to \bar tt$.\footnote{In the one-loop approximation, these decays are
  allowed, but only for $400\,\gev \simle M_A = M_{H^\pm} \simle 410\,\gev$
  and for $M_{H_2} \simge 450\,\gev$~\cite{Lane:2019dbc,Eichten:2021qbm}. The
  decays $H_2 \to W^\pm H^\mp$ and $ZA$ with two-loop-masses are discussed
  below.}

We focus on three types of BSM Higgs production and decay:

 \begin{itemize}

 \item[1.)] {\ul{$gg \to H^\pm \bar t b \to t \bar t b \bar b$ and $gg \to H_2 \to
       W^\pm H^\mp \to W^\pm \bar t b$}}

   There have been five searches for the first process relevant to the mass
   range of the GW-2HDM~\cite{Khachatryan:2015qxa,Aaboud:2018cwk,
     Sirunyan:2019arl, ATLAS:2021upq,CMS:2020imj}. The first of these was a
   CMS search at $8\,\tev$; the other four used $13\,\tev$
   data. Ref.~\cite{ATLAS:2021upq} is an ATLAS search using its full Run~2
   data set of $139\,\ifb$. Ref.~\cite{CMS:2020imj} is a CMS search using
   $35.9\,\ifb$ of $13\,\tev$ data taken in 2016; it is distinguished by
   having looked for the $\bar t b t\bar b$ final state in the all-jet mode.

   The $8\,\tev$ search by CMS ~\cite{Khachatryan:2015qxa} was used in
   Ref.~\cite{Lane:2018ycs} to set the limit $\tan\beta \simle 0.50$ for
   $180\,\gev < M_{H^\pm} \simle 500\,\gev$. The searches at $13\,\tev$ have
   not improved on this limit despite the larger data sets and, indeed, they
   have worse sensitivity at $M_{H^\pm} = 200$--$500\,\gev$ than the CMS
   8-TeV result. For example, the limit on $\tan\beta$ for
   $M_{H^\pm} = 200$--$500\,\gev$ extracted from the ATLAS $139\,\ifb$
   data~\cite{ATLAS:2021upq} is
   $\tan\beta < 1.10 \pm 0.14$~\cite{Eichten:2021qbm}. The reason for this
   disappointing outcome is the large $t \bar t$ background at low masses and
   the fact that it increases with collider energy faster than the signal.

   Given the payoff a significant improvement in the limit on $\tan\beta$ at
   low $M_{H^\pm}$ might have, we strongly urge ATLAS and CMS to find a way
   to improve the signal efficiency of this search. One possibility may be to
   use
\be\label{eq:H2WHpm}
gg \to H_2 \to W^\pm H^\mp.
\ee
Since $H^+$ decays to $t\bar b$, the final state in this mode,
$W^+ W^- b \bar b$, is the same as the near-threshold process above. But,
because it occurs at a higher invariant mass, kinematic cuts taking advantage
of that may provide a better signal-to-background ratio. The $H_2$ decay rate
is proportional to $p^3_W$ and, therefore, is sensitive to the available
phase space. It quickly becomes dominant when $M_{H_2}\simge 400\,\gev$ and
the $W$ is longitudinally polarized.\footnote{Decays such as this one were
  discussed in the one-loop approximation of the GW-2HDM in
  Refs.~\cite{Lane:2019dbc, Brooijmans:2020yij,Eichten:2021qbm}.} In the
perturbative calculation of the two-loop BSM masses practically the entire
allowed range of $M_{H^\pm}$ is covered --- from $180\,\gev$ to $365\,\gev$,
with $M_{H_2}$ ranging from $540\,\gev$ up to $700\,\gev$ and back down to
$510\,\gev$. In the amoeba method, the allowed region is restricted to
$M_{H^\pm} = 300$--$350\,\gev$, with $M_{H_2} = 525\,\gev$ down to
$450\,\gev$.\footnote{To our knowledge, this search has not been carried out;
  nor has one for $H_2 \to ZA \to \ellp\ellm t \bar t$. This decay has a more
  restricted allowed range; it is discussed below in item~3.}

\item[2.)] {\ul {$gg \to A/H_2 \to t \bar t$ and $gg \to t\bar t \to t\bar t
      A/H_2 \to t\bar t t\bar t$}}\hfil

  A search by CMS with $35.9\,\ifb$ of data at $13\,\tev$ for
  $\varphi = A/H_2 \to \bar tt$ with low mass, $400 < M_{A/H_2} < 750\,\gev$,
  is in Ref.~\cite{Sirunyan:2019wph}. Gluon fusion production proceeds
  through a top loop, and the principal background is $gg \to \bar tt$ near
  threshold. CMS presented model-independent constraints on the ``coupling
  strength''~$g_{\varphi t\bar t} = \lambda_{\varphi t\bar t}/(M_t/v)$ and
  for width-to-mass ratios $\Gamma_\varphi/M_\varphi = 0.5$--25\%. In the
  GW-2HDM, $g_{\varphi t\bar t} = \tan\beta$. For the $C\!P$-odd case,
  $\varphi = A$, with $400\,\gev < M_A < 500\,\gev$ and all $\Gamma_A/M_A$
  considered, the region $\tan\beta < 0.50$ was not excluded.\footnote{The same
    appears to be true for $\varphi = H_2$ with
    $\Gamma_{H_2}/M_{H_2} \simge 1\%$.} This is possibly due to an excess at
  $400\,\gev$ that corresponds to a global (local) significance of
  $1.9\,\,(3.5\pm 0.3)\,\sigma$ for $\Gamma_A/M_A \simeq 4\%$. The CMS paper
  noted that higher-order electroweak corrections to SM $gg \to t\bar t$
  threshold production may account for the excess and that further
  improvement in the theoretical description was needed.

  To ameliorate the effects of interference of the
  $gg \to A/H_2 \to t \bar t$ signal with SM $t \bar t$ production,
  CMS~\cite{CMS:2019rvj} and ATLAS~\cite{ATLAS:2022ohr} searched for
  $gg\to t \bar t$ with $A/H_2$ radiated from one of the top-quarks and
  decaying to $t\bar t$. Both experiments used their full Run~2 data sets,
  $137\,\ifb$ and $139\,\ifb$. For these data sets, the interference with SM
  four-top production was stated to be negligible. In this approach the
  experiments searched for a resonant $t \bar t$ excess in the four-top-quark
  data. They expressed $95\%$~CL upper limits on the signal cross section
  times $B(A/H_2 \to t\bar t)$ in terms of the type-II 2HDM of
  Ref.~\cite{Branco:2011iw}. In that model, the coupling of $A$ and $H_2$ is
  proportional $M_t\cot\beta/v$, and the experiments converted the
  $\sigma\cdot B$ limits into lower limits on $\tan\beta$. In the GW-2HDM,
  these translate into upper limits on $\tan\beta$.\footnote{See the note
    below Eq.~(\ref{eq:Z2}) in Sec.~I.} For CMS they are
  $\tan\beta < 1.6\,\,(0.7)$ assuming $M_A = M_{H_2} = 400\,\gev$ (assuming
  only $H_2$ with $M_{H_2} = 600\,\gev$); for ATLAS, they are
  $\tan\beta < 1.7\,\,(0.9)$ for $M_A = M_{H_2} = 400\,\gev$
  ($M_{H_2} = 600\,\gev$). These limits are much weaker than
  $\tan\beta < 0.50$ from the earlier CMS and ATLAS searches for
  $gg \to H^\pm t \bar b$. On the other hand, these four-top searches for a
  relatively low-mass $A$ or $H_2$ may benefit substantially from the High
  Luminosity LHC.

\item[3.)] {\ul{$gg \to H_2 \to ZA$}}

  There have been three published searches for $H_2 \to ZA$ with
  $ZA \to\ellp\ellm \bar bb$, where $\ell = e$ or $\mu$:
  \cite{Aaboud:2018eoy,CMS:2019ogx,ATLAS:2020gxx}. The latter ATLAS search
  updated the former one with the full Run~2 data set. As with
  $gg \to H_2 \to W^\pm H^\mp$, these $H_2$ decay rates are proportional to
  $p^3_Z$. They were discussed in the one-loop approximation and two
  comparisons to the GW-2HDM were presented in Refs.~\cite{Lane:2019dbc,
    Eichten:2021qbm}. Two examples were presented, one of which (with
  $M_{H_2} = 500\,\gev$, $M_A = 300\,\gev$ and $\tan\beta = 0.50$) was
  excluded at the $95\%$~CL in the newer ATLAS search.

  Another approach, without the large $b\bar b$ background, is to use
  $A \to t\bar t$. In the two-loop perturbative method, the region
  $M_A = 350$--$365\,\gev$ corresponds to $M_{H_2} = 630\,\gev$ down to
  $508\,\gev$ and has substantial ($> 20\%$) branching ratios of
  $H_2 \to ZA$. In the amoeba method, there is no $M_{H_2} > M_A + 100\,\tev$
  for which $M_A > 2M_t$. It's worth a try; nothing ventured, nothing gained.

\end{itemize}

\vfil\eject 


\section*{Acknowledgments} EE thanks the Fermi Research Alliance, LLC under
Contract No.~DE-AC02-07CH11359 with the U.~S.~Department of Energy, Office of
Science, Office of High Energy Physics. KL thanks Laboratoire
d'Annecy-le-Vieux de Physique Th\'eorique (LAPTh) for its hospitality during
the final stage of this work. We are very grateful for valuable guidance from
Stephen Martin. We also thank Kevin Black, Jon Butterworth, Alvaro De~Rujula,
Ulrich Heintz, Guoan Hu, and William John Murray for many helpful
discussions and suggestions.


\bibliography{Two-Loop}
\bibliographystyle{utcaps}
\end{document}